\documentclass[aip,pop,12pt,preprint]{revtex4-1}
\pdfoutput=1
\usepackage{graphicx}
\def\lde{\lambda_{De}}
\def\lesim{\ \hbox to 0 pt{\raise .6ex\hbox{$<$}\hss} \lower.5ex\hbox{$\sim$}\ }
\def\gesim{\ \hbox to 0 pt{\raise .6ex\hbox{$>$}\hss} \lower.5ex\hbox{$\sim$}\ }
\begin{document}
\title{Collisional Effects on Nonlinear Ion Drag Force for Small Grains}
\author{I H Hutchinson} \author{C B Haakonsen} \affiliation{Plasma
  Science and Fusion Center and\\ Department of Nuclear Science and
  Engineering,\\ Massachusetts Institute of Technology,\\ Cambridge,
  MA, USA.}

\begin{abstract}
  The ion drag force arising from plasma flow past an embedded
  spherical grain is calculated self-consistently and
  non-linearly using particle in cell codes, accounting for
  ion-neutral collisions. Using ion velocity distribution appropriate
  for ion drift driven by a force field gives wake potential and
  force greatly different from a shifted Maxwellian distribution,
  regardless of collisionality. The low-collisionality forces are
  shown to be consistent with estimates based upon cross-sections for
  scattering in a Yukawa (shielded) grain field, but only if
  non-linear shielding length is used. Finite collisionality initially
  enhances the drag force, but only by up to a factor of 2. Larger
  collisionality eventually reduces the drag force. In the collisional
  regime, the drift distribution gives larger drag than the shift
  distribution even at velocities where their collisionless drags are
  equal. Comprehensive practical analytic formulas for force that fit
  the calculations are provided.
\end{abstract}

\maketitle

\section{Introduction}

The drag force arising from the interaction of flowing ions with a
negatively charged grain embedded in a plasma has been the subject of
intensive study during the recent past. It is a problem of substantial
intrinsic interest as well the basis for understanding much of the
behavior of grain equilibrium and dynamics in dusty plasmas. Yet
reliable values for the drag in collisional plasmas are
still not known.

Neglecting ion-neutral and ion-electron collisions, the classic
Coulomb orbit impact parameter integral
treatment\cite{chandrasekhar43,Cohen1950,Rosenbluth1957}, used for the
plasma particles themselves, can be applied to a grain when the plasma
electron Debye length ($\lde$) is sufficiently large compared with the
grain radius ($r_p$). Strictly speaking, it is when $\lde$ is much
greater than the 90-degree scattering impact parameter for a typical
ion velocity ($v_i$),
\begin{equation}
  \label{eq:b90}
  b_{90} \equiv {Ze|Q|\over 4\pi \epsilon_0 m_i v_i^2}, 
\end{equation}
where $m_i$, $Z$ are the ion mass and charge-number, and $Q$ is the
charge on the grain. However, for dust grains (unlike elementary
particles) $r_p$ and hence $Q$ are usually large enough that
$b_{90}/\lde$ is not small, and so the standard treatment is
inapplicable, because the argument of the Coulomb logarithm is not
large. For example if $Z=1$, $\lde\sim 600\mu$m, $T_e/T_i\sim100$ and
grain potential is $\phi_p \sim -2 T_e/e$, representing a typical
dusty plasma experiment, then at the ion thermal velocity,
$b_{90}/\lde \sim 2 (r_p/\lde)(T_e/T_i) \sim r_p/3\mu $m. In such a
plasma, even for grains at the lower limit of being able to be
detected by unperturbative laser illumination (optimistically perhaps
$r_p\sim 0.3\mu$m), $b_{90}/\lde$ is too large for the cut-off Coulomb
logarithm to give accurate results\cite{Kilgore1993}. Practically all
dusty plasma experiments with approximately room-temperature ions are in the
non-linear regime $b_{90}/\lde \gesim 1$.

Ions moving with superthermal directed velocity exhibit smaller
$b_{90}$. For example, at the sound speed a value $b_{90}/\lde \sim 2
(r_p/\lde)$ results (smaller by $T_i/T_e$), which would not usually
imply strong nonlinearity. However, as we shall see in a moment, the
correct distribution function for ions whose drift is driven by
electric field against drag arising from neutral collisions, is
\emph{not} a shifted Maxwellian at the flow speed. Instead the
distribution retains a substantial fraction of its population at low
ion speed, and hence still in the non-linear regime $b_{90}/\lde \gesim 1$.

If the shielded potential of the grain is approximated by a
Debye-H\"uckel (Yukawa) form $\phi = \phi_p
\exp[-(r-r_p)/\lambda]r_p/r$, then, even into the nonlinear regime
($b_{90}/\lde > 1$), one can calculate the momentum transfer
cross-section as a function of ion velocity by integrating (generally
numerically) ion
orbits\cite{Liboff1959,kihara59,mason67,hahn71,Khrapak2002}. However,
even if this potential remains valid, despite the inadequacy of the
linearized Boltzmann approximation on which the Yukawa form is based,
it is unclear exactly what value to use for the shielding length
($\lambda$) even though some important
studies\cite{Daugherty1992,Choi1994} address the question. This is
especially difficult for $T_e/T_i \gg 1$, since the ions then dominate
the shielding length. Moreover, direct absorption of ions that strike
the grain must be accounted for.

The self-consistent ion drag force on a single grain in a
\emph{collisionless} plasma has previously been determined from
particle-in-cell calculations\cite{Hutchinson2005,hutchinson06} (using
the code SCEPTIC), as a function of the flow velocity $v_f$ of a
shifted Maxwellian ion distribution, over a range of values of
$\lde/r_p$, $T_e/T_i$, for electrically floating grains. The values
differ significantly (up to factor of 2) from the best binary
collision calculations\cite{Khrapak2002,Khrapak2003,khrapak05}. The
differences arise because the shielded potential form assumed in those
theories is not what actually occurs. There are important
spherical-asymmetries in the ion density and potential that are not
accounted for by the heuristic potential assumptions. A practical
numerical expression\cite{hutchinson06}, which approximates the PIC
calculations, covers most of the relevant parameter space for
collisionless plasmas with shifted Maxwellian ions.

A major shortcoming of collisionless drag force calculations is that
many experiments are carried out in regimes where ion-neutral
collisions are important. The neutral density in many dusty plasma
experiments is such that charge-exchange collisions between ions and
neutrals can occur with mean-free-path less than $\lde$. For example, 
in argon gas pressures between 10 and 100 Pa, the typical collision
mean free path is in the approximate range 1mm to 0.1mm. Collisionless
treatments are therefore suspect, and the question arises as to what
effects collisions have. 

Actually, even for very low collisionalities the collisions can be
important. The most important effect is that the velocity distribution
of ions is not a shifted Maxwellian in a situation where the drift is
driven by a force field (for example background electric field) in a
\emph{stationary neutral}
background\cite{Fahr1967,Rebentrost1972,Ivlev2005,Patacchini2008,Lampe2012}. The
change of the distribution function form is independent of the level
of collisionality. It occurs formally even in the low-collisionality
limit. For ion drift that exceeds the neutral thermal speed (very
often the case) the resulting distribution retains, for example, a
substantial fraction of the ions with velocities near the (neutral)
thermal velocity $v_i\sim v_{th}$, which is not the case with a
shifted Maxwellian. This difference of ion distribution alone has
important effects on the ion drag force.

Taking into account a finite level of collisionality brings into play
effects that have so far defied incorporation into the binary
collision treatment. As an alternative, kinetic-theory calculations
that utilize the linearized plasma response\cite{Neufeld1955} have
been pursued\cite{Ivlev2004,Ivlev2005}, which can incorporate both the
non-Maxwellian background ion distribution, and the direct effects of
collisions. Unfortunately these calculations confront the fact that
most experimental dusty plasma parameters are not such as to justify
the linear approximation. The wave-number integrations that the
linearized response utilizes must be heuristically cut off to prevent
unphysical divergences, and the argument of the resulting logarithmic
factor is $\sim \lde/b_{90}$: not large. The quantitative results are
therefore unreliable. 

Again some SCEPTIC calculations\cite{Patacchini2008} have given fully
self-consistent values of the force incorporating both the correct
background distribution and the direct effects of collisions. These
results have contradicted some of the speculations about
self-acceleration of dust grains, showing that while it is indeed
possible for the ion drag force to reverse direction, it does so only
deep into the strongly collisional (short-mean-free-path) regime where
other forces on the grain are dominant. However, these simulations
have so far been limited to sound-speed level drift velocities, and no
comprehensive practical fit to the collisional drag force has yet been
developed.

In the mean time, the wake potential
structure\cite{lampe00,lampe05,Hutchinson2011a,Dewar2012,Ludwig2012} 
 has
been under investigation in order better to understand the mutual
interaction of multiple
grains\cite{Hutchinson2011b,HutchinsonLowM2012}. Most such
calculations have so far used shifted Maxwellian ion
distributions. However, the non-Maxwellian character expected with
even vanishingly small level of collisionality has a major effect on
the wake structure because it completely changes (generally greatly
enhances) the Landau damping which is usually responsible for the
decay of the potential oscillations.

The purpose of the present article is to provide a wide ranging
numerical exploration of the consequences of charge-exchange neutral
collisions for the ion drag on a small grain in a uniform,
drifting-ion plasma, accounting fully for the non-linearity of the
problem. The effects of the ion distribution function shape changes
are documented separately from the
collisionality itself, and relatively small flow velocities as well
as sonic and supersonic investigated. Because of the rather large
number of relevant parameters even ignoring any magnetic field (as we
do here) the coverage of the parameter space is limited. In particular
for our comprehensive investigations we treat only the case
$T_e/T_i=100$, which is the upper end of the applicable temperature
ratio range, and we choose to fix the grain potential to
$\phi_p=-2T_e/e$, a typical floating value.

Section \ref{sec2} explains the computational techniques and outlines
the analytic treatments that are to be compared with the simulations.
Section \ref{sec3} compares the plasma shielding/wake character
observed in computational simulations for the shifted Maxwellian and
driven-drift non-Maxwellian background ion distributions. It also
documents the ion drag force for negligible level of collisionality
and demonstrates the sufficiency of the analytic collisionless
`fits'. In engineering parlance these fits are empirical
`correlations' though motivated by the underlying physics. Section
\ref{sec4} contains the numerical simulation results for a wide range
of collisionality, developing a practical fit to express the influence
of collisions as a correction factor for the ion drag force relative
to the collisionless value. Section \ref{sec5} offers a heuristic
explanation of the calculated trends with collisionality.

\section{Techniques, Theory, and Collisionless Fits}
\label{sec2}

\subsection{Particles in Cell}

The particle in cell computational approach to representing the
interaction of the flowing plasma with a spherical grain has been
described in detail elsewhere\cite{Hutchinson2002,Hutchinson2003}.  A
large number of ions are moved in six-dimensional phase space, under
the influence of the electric field arising from the self-consistently
calculated potential $\phi$, plus an optional force ${\bf D}$, to be
explained in the next subsection:
\begin{equation}
  \label{eq:ionmotion}
  m_i {d{\bf v}\over dt} = -Z e \nabla \phi + {\bf D} .
\end{equation}
The potential is
represented on a cellular grid and satisfies
\begin{equation}
  \label{eq:poisson}
  \nabla^2 \phi = {e\over \epsilon_0}(Z n_i - n_e).
\end{equation}
The ion density ($n_i$) is determined by depositing the individual
ions onto the grid (using Cloud in Cell\cite{*[{}][{, p20ff}] birdsall91} particle
shape) and the electron density is assumed to be governed by a thermal
Boltzmann factor
\begin{equation}
  \label{eq:boltzmann}
  n_e = n_{e\infty} \exp(e\phi/T_e).
\end{equation}
Thus the treatment can be considered ``hybrid'' particle in cell and
does not need to resolve the electron plasma time-scale.

Ions fill a computational region across whose outer boundary they
leave or enter. Those entering are injected in accordance with the
distribution function presumed for the external, uniform, plasma.  The
grain is represented by a fixed-potential sphere embedded in the
computational region. It absorbs ions that strike its surface, but
it does not emit any ions. 

The conditions on potential at the outer computational boundary are
chosen to represent most conveniently a potential decaying to zero at
infinity ($d\ln\phi/d\ln r =-1$). Trials show that the precise choice of
distant boundary condition does not affect the drag force.

The majority of the calculations here are carried out with the SCEPTIC
code\cite{Hutchinson2002,Hutchinson2003}, which uses a spherical
cell-grid conformal with the grain surface. The radial grid is
uniform, which means that the cell-volume is proportional to the
square of the radius ($r$). Up to 400 radial cells are used, depending on the
domain size required to provide a converged result. The domain must
typically be at least $\sim2\lde$ in radius to give accurate forces.
Angular cells are equally spaced in $\cos\theta$ where $\theta$ is the
positional angle with respect to the drift direction ${\bf z}$; 100
angular cells are used. The potential is presumed 2-dimensional,
independent of the azimuthal angle. The resulting cell size is much
smaller than the Debye length, and sufficient to resolve the potential
variation to an accuracy of a few percent. Up to 32 million ions are
advanced for up to 4000 steps. A substantial fraction of those steps are in
steady state and so can be averaged-over to improve statistics.

Some comparison calculations are made with the COPTIC
code\cite{Hutchinson2011a} to verify computational
accuracy, confirm that domain size and shape are not substantially
affecting the results, and explore extended wakes. COPTIC uses a
cartesian grid but can accommodate oblique curved surfaces of embedded
objects\cite{HutchinsonCart2011}, such as the spherical grain that is used.

In either code, the force on the grain is calculated by accounting for
the total momentum flux transferred by ions, electron pressure, and
Maxwell stress inward across spherical surfaces containing the
grain\cite{Hutchinson2005}.  The volumetric collisional momentum loss
to neutrals within the sphere\cite{Patacchini2008} is subtracted and
what remains is the total momentum transfer (rate) to the grain, i.e.\
the drag force on it. The fact that in steady state the momentum
transfer is independent of the radius of the measurement sphere allows
one to establish good convergence of the calculations by measuring at
a number of different radii and observing equal results.

 Because we are interested in quite large ratios of
Debye length, $\lde$, (which constrains minimum domain size) to grain
radius, $r_p$, (which
constrains maximum grid spacing at the grain) we are challenged in
terms of computational capacity, and content ourselves with overall
uncertainty (both noise and spatial resolution error) of force
measurement of up to approximately 5\% (but usually smaller) plus an
absolute uncertainty of roughly $2r_p^2n_{e\infty}T_e$.

\subsection{Collisions and Distributions}

Charge-exchange collisions of ions with neutrals are represented in
the code as a replacement of the ion velocity with a new velocity drawn
randomly from the neutral velocity distribution. The simulations here
use a collision frequency that is independent of velocity (equivalent
to a BGK collision operator). Although not a perfect representation of
the actual collision cross-section, this is sufficient to demonstrate
the important new physics.

The nature of the collisions and the cause of the ion drift determine
the self-consistent background distribution. The simplest case is when
the ion drift is a reflection just of a neutral drift. In that case,
which incidentally is equivalent by Galilean transformation to
studying a grain moving through a stationary plasma, the ion velocity
distribution in the unperturbed uniform plasma is simply equal to the
neutral velocity distribution --- which is the birth distribution of
the ions after collisions. The extra force ${\bf D}$ is then zero.

The opposite limit is when neutrals are (on average) stationary and
the ion drift is driven by a uniform force field ${\bf D}$. The nature
of this force field is generally unimportant for the calculations
(except possibly in respect of the boundary conditions at the grain
surface). It is usually thought to be a background electric field, but
could theoretically equally well be gravity or some other unspecified
force. It is important that the Boltzmann factor for electrons not
take account of any potential gradient that is responsible for this
driving force field ${\bf D}$ because otherwise the solution of the
system in the absence of a perturbing grain would not be uniform. That
is why it is convenient to separate it out from the potential
arising because of the grain: $\phi$. 

Intermediate cases are possible where the drift is partly driven by
${\bf D}$ and partly by neutral drift. Their background ion
distribution can then be represented by a shift of the collisional
distribution. However, we present no simulation of such intermediate
cases here. Different assumptions about collision cross-section
dependence upon velocity lead to somewhat different distribution
shape\cite{Lampe2012}, they share many of the same features.

The solution of the ion Boltzmann equation for the unperturbed
background plasma when the neutrals are Maxwellian has two very
different forms, corresponding to shifted or stationary neutrals. We
call these the \emph{shift} distribution (${\bf D}=0$) and the
\emph{drift} distribution (${\bf D}\not=0$) respectively.  The shift ion
distribution is simply equal to the neutral distribution: a shifted
Maxwellian [$\exp(-m_i\{v-v_f\}^2/2T_n)$]. (The ion temperature can
then be considered simply equal to the neutral temperature $T_i=T_n$
and we will not hereafter draw any distinction between $T_i$ and
$T_n$).

\begin{figure}[htp]
  \centering
  \includegraphics[width=3.5in]{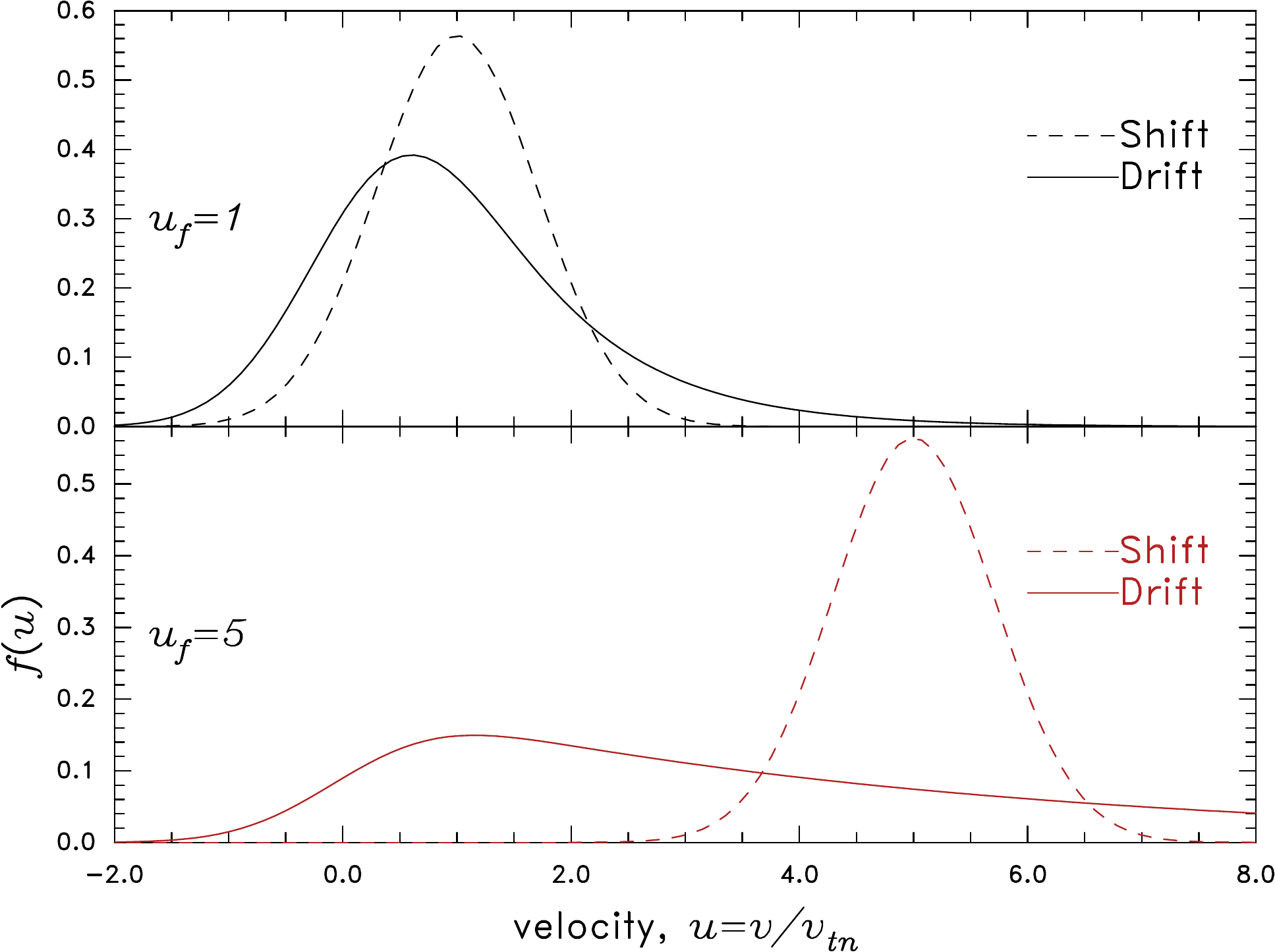}
  \caption{Shift (shifted Maxwellian) and Drift (eq.\ \ref{eq:driftdist}) ion distributions for two values of the flow velocity, plotted versus velocity $u$, normalized to the neutral thermal velocity.}
  \label{distribs}
\end{figure}

By contrast, in the absence of neutral shift, the drift solution of the
Boltzmann equation for the ions gives a distribution that is the
convolution of a stationary Maxwellian [$\exp(-m_iv^2/2T_n)$] with a
half-exponential function representing the Poisson collision
statistics [$\exp(-v_z/v_f)$ for $v_z>0$, $z$ being the direction of
${\bf D}$]. The resulting ion distribution can be expressed analytically
as\cite{Patacchini2008}:
\begin{equation}
  \label{eq:driftdist}
  f_i({\bf v}) = {n_i\over \pi v_{tn}^2}{1\over 2v_f} \exp\left(-{v^2\over v_{tn}^2}\right) {\rm erfcx}
\left({v_{tn}\over 2 v_f} - {v_z\over v_{tn}}\right) 
\end{equation}
where $v_{tn}=\sqrt{2T_n/m_i}$ is the thermal speed and ${\rm
  erfcx}(x) \equiv \exp(x^2){\rm erfc}(x)$. 
Notice that this form does
not depend upon the actual value of the collision frequency $\nu_c$ or
${\bf D}$, only the mean ion velocity $v_f$ and the thermal speed. Of
course there is a definite relationship,
\begin{equation}
  \label{eq:colrel}
  v_f = {D\over m_i \nu_c},
\end{equation}
between ${\bf D}$ and $\nu_c$. 
Examples of the shifted Maxwellian distribution and the drift
distribution are shown in Fig.\ \ref{distribs} 
When $v_f\gg v_{tn}$ the ion
distribution in the direction $\hat{\bf z}$, $f_z(v_z) = \int\int f_i
dv_x dv_y$ is approximately a half-exponential $f_z(v_z) \approx n_i
\exp(-v_z/v_f)/v_f$ for $v_z>0$. An important feature is that its
magnitude at small $v$ is just $n_i/v_f$, inversely proportional to
$v_f$, rather than falling rapidly to zero as it does for a shifted
Maxwellian. This feature is common to any ${\bf D}$-driven drift
distribution, not just to the constant-$\nu$ approximation used here.

\subsection{Yukawa drag force}

It is helpful to compare the PIC calculations at low collisionality,
with estimates based upon previously published momentum-transfer
cross-sections calculated from orbits in Yukawa form potentials $\phi
= Q\exp(-r/\lambda)/(4\pi \epsilon_0 r)$. Writing
$\beta=b_{90}/\lambda = ZeQ/(4\pi \epsilon_0 m_iv^2 \lambda)$, a fit
to the point-charge cross-section in the low-beta region $\beta < 1$,
has been given as\cite{khrapak04submit}
\begin{equation}
  \label{eq:sigma1}
  \sigma_1 = \pi \lambda^2\beta^2 4\ln(1+1/\beta) = \pi b_{90}^24\ln(1+1/\beta),
\end{equation}
In the limit of low $\beta$
these become the classic cut-off Coulomb drag.  In the high-beta
region, $\beta\gg 1$, an asymptotic form is\cite{khrapak04submit}
\begin{equation}
  \label{eq:sigma2}
  \sigma_2 = \pi\lambda^2
  0.8[\ln^2\beta+2\ln\beta+2.5
]. 
\end{equation}
Although not perfect near $\beta=8$, a reasonable fit to the
Yukawa cross-section values, valid over the entire range, and having
the correct limits, is to use for the combined cross-section,
$\sigma_y$, a weighted sum of inverses thus: 
\begin{equation}
  \label{eq:sigmas}
  \sigma_y^{-1} =  [\sigma_1^{-1} + (0.15\beta)^2
    \sigma_2^{-1} ]/[1+(0.15\beta)^2] . 
\end{equation}

To obtain the total drag force on a spherical grain of finite radius
$r_p$, it has been shown\cite{Khrapak2002,khrapak05,hutchinson06} that a good
prescription is to change the form of the logarithm term in eq.\
(\ref{eq:sigma1}) to
\begin{equation}
  \label{logadjust}
  \ln\Lambda = \ln \left( b_{90}+ \lambda\over  b_{90}+r_p\right)
= \ln \left(\beta +1  \over \beta+ r_p/\lambda \right)
\end{equation}
and add the momentum transfer by direct collection of those
ions whose impact parameter is smaller than the OML critical impact
parameter
\begin{equation}
  \label{eq:bcrit}
  b_c = r_p \sqrt{1+ e|\phi_p|2/m_iv^2},
\end{equation}
so that the force (in the $z$-direction) from this Yukawa-potential
treatment is\footnote{For very high values of $\beta$, the collection
  radius can be no larger than a few $\lambda$ or else the OML
  validity conditions are
  violated\protect{\cite{khrapak04submit}}. This limitation proves to
  be unimportant for the high flow velocities at which the direct
  collection is an important fraction of the force.}
\begin{equation}
  \label{eq:binaryforce}
  F_Y = F_o + F_c = \int (\sigma_y + \pi b_c^2) m_i v_z v f_i({\bf v}) d^3v.
\end{equation}
Because we have from our PIC codes convenient representations of
$f_i$, we can evaluate this integral numerically when comparing the PIC
evaluations of the force with the binary collision treatment.  The
comparison is then not affected by additional approximations.  We
note, however, that there are already both significant approximations
in using a Yukawa form, and significant uncertainties as to what value
of $\lambda$ to adopt.

\subsection{Shielding length}

It is often presumed that the shielding length to be used in
collisional calculations of ion drag force is given by the so called
linearized shielding length for which
\begin{equation}
  \label{eq:shieldlinear}
  \lambda^{-2} = \lde^{-2} + \lambda_{Di}^{-2} .
\end{equation}
Because $\lambda_D^{-2} = (Z^2 e^2 n/\epsilon_0 T)$ and $T_i\ll T_e$, the ion
Debye length generally dominates in this expression unless the
effective ion temperature is enhanced by sonic-speed-level ion flows.

However a Yukawa form potential with this shielding length is the
solution to a linear approximation $n_i/n_{i\infty} = 1-Ze\phi/T_i$
which is totally unjustified when $b_{90}/\lambda \gesim 1$, i.e.\ in
the nonlinear regime. Although some studies\cite{Daugherty1992} have
indicated that a Yukawa potential profile with a shielding length like
this may not be too far wrong, those studies used mono-energetic ions,
which is obviously a serious misrepresentation of the likely ion
distribution function even when the drift velocity is small. Actually
there exist analytic expressions for the spherically symmetric ion
density in the vicinity of an ion-attracting probe in a stationary
Maxwellian plasma\cite{Alpert1965}.  When intermediate energy barriers
are ignored they reduce in the (applicable) limit $\lambda/r_p,\
|Ze\phi|/T_i\gg 1$ to
\begin{equation}
  \label{eq:attracted}
  n_i = n_{i\infty} \sqrt{-4Ze\phi/\pi T_i}.
\end{equation}
In other words, the ion-density rises proportional to the square root
of $|\phi|$, not proportional to $|\phi|$. We will show that the nonlinear eq.\
(\ref{eq:attracted}) agrees reasonably well with the ion density that
SCEPTIC finds. 

Unfortunately no ready analytic expression is available to solve the
nonlinear Poisson equation that then arises. And numerical solutions show
poor resemblance to the Yukawa form. However, a simple cut-off
approximation can provide an appropriate scaling when $\lambda\gg
r_p$ and ions dominate shielding, as follows. Suppose that the ion density is
given by eq.\ (\ref{eq:attracted}). Suppose also that the dominant
radial dependence of $\phi$ is the $1/r$ Coulombic variation, but only
out to a cutoff radius. Take that cutoff radius $r_c$ to be the place
where the total ion charge within it is equal to $-Q/2$, so as to shield
half the grain field. Then one can readily solve for $r_c$ and find
\begin{equation}
  \label{eq:rcutoff}
  r_c = \left(|Q|e\over 4\pi \epsilon_0 T_e r_p\right)^{1\over5}
  \left(5\sqrt{\pi}\over 8\right)^{2\over5} 
\left({r_p\over \lde}{T_i\over T_e}\right)^{1\over5} \lde 
  \approx 1.2 \left({r_p\over \lde}{T_i\over T_e}\right)^{1\over5} \lde ,
\end{equation}
where the second form takes the grain potential to be
$-2T_e/e$. Applying this scaling to a Yukawa cross-section is purely
heuristic, but it gives a significantly longer shielding length.
Table \ref{shieldlengths} illustrates the values of nonlinear
shielding length, when $T_e/T_i=100$. The corresponding linear values
are all $\lambda/\lambda_{De}=0.1$, $\lambda/r_p=0.1\lambda_{De}/r_p$.
\begin{table}[htp]
  \centering
  \begin{tabular}{|l|c|c|c|c|c|c|}
    \hline
     & $\lambda_{De}/r_p:$ & 10 & 20 & 50 & 100 & 200
    \\ 
    \hline
    NonLinear & $r_c/\lambda_{De}$ (eq \ref{eq:rcutoff}) & 0.30 & 0.26
    & 0.22 & 0.19 & 0.17 \\
    & $r_c/r_p$ & 3.0 & 5.2 & 10.9 & 19 & 33 \\
  \hline
  \end{tabular}
  \caption{Nonlinear shielding lengths.}
  \label{shieldlengths}
\end{table}

Of course, an overall ion flow increases the effective shielding
length even more, by reducing ion shielding through increased ion
energy. For a shift distribution, it was found\cite{hutchinson06} that
appropriate shielding was obtained at substantial $v_f$ using an
almost-linear form
\begin{equation}\label{eq:almostlin}
\lambda_\ell^2 = r_p^2+\lde^2/[1+ZT_e/(T_i+ {\cal E})],
\end{equation}
where
\begin{equation}
  \label{adjust}
  {\cal E}= {\cal E}_s={0.5}m_iv_f^2\left[1+|v_f/0.4c_s|^3\right]
\end{equation}
represents the effects of flow.\footnote{Actually in
  \cite{hutchinson06} the denominator of the cubic term consisted of
  three terms to accommodate floating potential variation due to ion
  mass changes, and ion temperature variation: $[0.6+0.05\ln(m/Z) +
  (\lde/5r_p)(\sqrt{T_i/ZT_e}-0.1)]c_s$. The simplification to the
  present expression improves the fit, especially in the vicinity of
  $v_f\sim 0.2 c_s$. We are not here investigating changes in grain
  potential.}

We find the nonlinearity at low $v_f$ is accommodated better by using
$r_c$ rather than $\lambda_{\ell}$ for the shielding length. A plausible
and preferable nonlinear interpolation instead of eq.\ (\ref{eq:almostlin}) is
\begin{equation}
  \label{eq:shiftshield}
  \lambda^{-2} =  \lde^{-2} +(r_c^2+\lde^2{\cal E}/T_e)^{-1}.
\end{equation}

However, for drift distribution, the decrease in shielding with flow
is anticipated to be predominantly by reducing the density of the
low-velocity ion component proportional to $n_i/v_f$. On this basis,
the effective ion Debye length for shielding when $v_f\gg v_{ti}$ is
obtained by replacing $n_i$ (in $\lambda_{Di}=\sqrt{\epsilon_0
  T_i/Z^2e^2 n_i}$) by $n_i v_{ti}/v_f$, which is equivalent to adopting
\begin{equation}\label{eq:linvd}
{\cal E}={\cal E}_d=T_e\sqrt{T_i/ZT_e}\;v_f/c_s,
\end{equation}
in which the drift effect enters linearly in $v_f$ rather than
quadratically as in ${\cal E}_s$.
 The precise coefficient of $v_f$ in
eq.\ (\ref{eq:linvd}) is of course uncertain, but taking this
simple one gives good results, when substituted into eq.\
(\ref{eq:shiftshield}), as we shall see.

\subsection{Collection force expressions}

For future use, it is convenient to have closed form
expressions for the drag force eq.\ (\ref{eq:binaryforce}) that are
easy to evaluate without having to do integrals. We present these
here and in the next subsection.

The $b_c^2$ term of eq. (\ref{eq:binaryforce}) is from direct ion
collection. It does not depend upon the shielding form in the OML
limit. For a shifted Maxwellian distribution, it can be integrated
directly to yield\cite{Hutchinson2005}:
\begin{eqnarray}\label{Fc}
F_{cs}(u_f) &=& n_i r_p^2 T_i \sqrt{\pi}
\Big\{u_f(2u_f^2+1+2\chi){\rm e}^{-u_f^2} +\\
&&\left[4u_f^4 + 4u_f^2 -1 -2(1-2u_f^2)\chi\right]
{\textstyle{\sqrt{\pi}\over 2}}{\rm erf}(u_f)
\Big\}/u_f^2\ ,\nonumber
\end{eqnarray}
where $u\equiv v/v_{ti}$ denotes velocity normalized to ion thermal
velocity and  $\chi\equiv -Ze\phi_p/T_i$ is potential normalized to
ion temperature.

\begin{figure}[htp]
  \centering
  \includegraphics[height=2.6in]{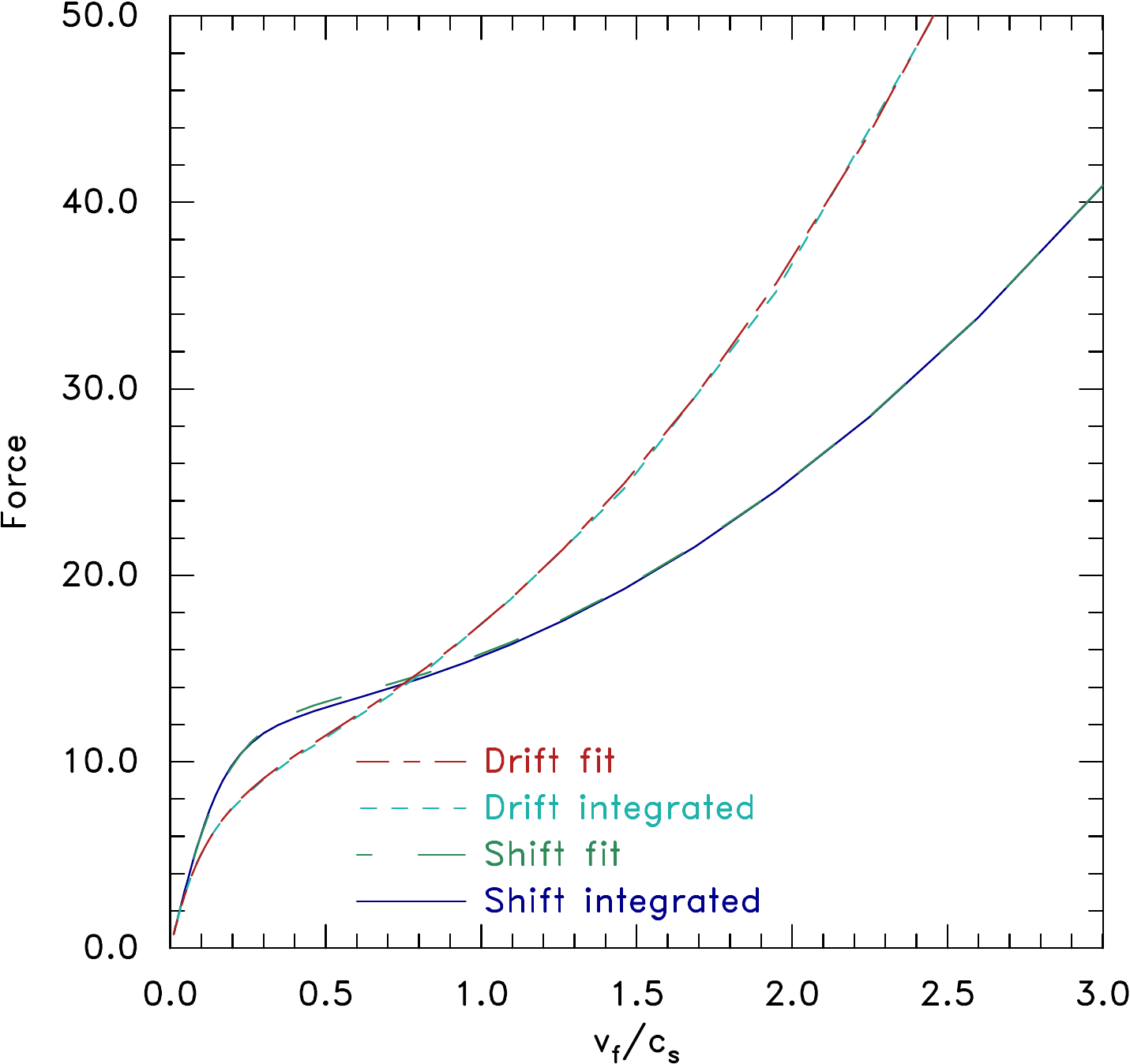}
  \caption{Direct ion collection force analytic
    fits, eqs.\ (\ref{Fcapprox}) and (\ref{FcdFinal}) compared with numerical integration. $\chi=200$.}
  \label{colfrc}
\end{figure}

To develop analytic expressions for the \emph{drift} distribution, we
recall that the drift distribution function can be written as the
convolution of the Maxwellian (neutral) distribution with a
half-exponential:
\begin{equation}
  \label{convol}
  f({\bf v}) = {1\over u_f} \int_0^\infty {n_i\over v_{ti}^3\pi^{3/2}}
    e^{-({\bf u}-s{\bf \hat{z}})^2} e^{-s/u_f} ds.
\end{equation}
Since we have the closed form expression (\ref{Fc}) for the collection
force of a shifted Maxwellian distribution ($e^{-({\bf u}-s{\bf \hat{z}})^2}$) already,
we can therefore write immediately the drift distribution collection force as
\begin{equation}
  \label{Fcd}
  F_{cd}(u_f) = {1\over u_f} \int_0^\infty F_{cs}(s) e^{-s/u_f} ds.
\end{equation}
Unfortunately eq.\ (\ref{Fc}) does not lead to an integral that can
conveniently be evaluated in eq.\ (\ref{Fcd}). However, an
accurate ($\sim 2\%$) approximation to eq.\ (\ref{Fc}) is 
\begin{equation}
  \label{Fcapprox}
  F_{cs}(u_f) = n_i T_i r_p^2 2 \pi \{u_f^2 +(1+\chi)[1-(1+bu_f)e^{-au_f}]\}
\end{equation}
where $b=0.8$ and
\begin{equation}
  \label{aeq}
  a = b+ {(16+8\chi) \over 6\sqrt{\pi}(1+\chi)}.
\end{equation}
The basis of this approximation is that for large $u_f$ the exponential
term is negligible and the remainder agrees with the asymptotic limit
of eq.\ (\ref{Fc}). At $u_f\to 0$, the form of $a$ is chosen to give the
correct linear dependence. And finally $b$ is chosen ad hoc to improve
the fit in the intermediate-$u$ region.  Fig.\ \ref{colfrc} shows that
eqs.\ (\ref{Fcapprox}) and (\ref{Fc}) are practically
indistinguishable.

Substituting eq.\ (\ref{Fcapprox}) with argument $s$ in place of
$u_f$, into eq.\ (\ref{Fcd}) gives an integral that can be evaluated
as the following elementary expression:
\begin{equation}
  \label{FcdFinal}
   F_{cd}(u_f) = n_i r_p^2 T_i 2 \pi \left[ 2 u_f^2 + (1+\chi) {(a-b)u_f +
     (au_f)^2 \over (1+au_f)^2}\right],
\end{equation}
which is again indistinguishable from the numerically integrated result.

Equations (\ref{Fcapprox}) and (\ref{FcdFinal}) provide accurate and
easily evaluated expressions for the direct collection ion force with
shift and drift distributions respectively, but of course only when
the actual level of collisionality is negligible. Notice that the
$u_f^2$ term, dominant at large flow, is twice as large for the drift
distribution as for the shift distribution.

\subsection{Orbital Drag Force Expressions}
\label{orbitalsub}

The orbital part of the collisionless ion drag can be expressed
as\cite{hutchinson06}
\begin{equation}
  \label{dragshift}
  F_o =  n_e T_e r_p^2 \left(Ze\phi_p\over T_e\right)^2 {T_e\over T_i} 4\pi G(u_f) \ln\Lambda,
\end{equation}
which we will show compares favorably with the computational results.

For the \emph{shift} distribution the function $G(u_f)$ is simply the
Chandrasekhar function
\begin{equation}
  \label{Chandra}
G_s(u)\equiv \left[{\rm erf}(u) - 2u {\rm
e}^{-u^2}/\sqrt{\pi}\right]/(2u^2)\ .  
\end{equation}
The logarithm is written using the almost linear screening length, eq.\
(\ref{eq:almostlin}), as
\begin{equation}
  \label{coullog}
\ln\Lambda_s=\ln\left( b_{90}+ \lambda_\ell\over  b_{90}+r_p\right),  
\end{equation}
and
\begin{equation}
  \label{b90s}
  b_{90s}= Ze\phi_p/(2T_i+{\cal E}_s) ,
\end{equation}
where ${\cal E}_s$ is given by eq.\ (\ref{adjust}).
In effect, this is a slightly adjusted form of the fit given
in  \cite{hutchinson06}.

For the \emph{drift} case, because the relevant integrals cannot be performed
analytically in the nonlinear regime, ad hoc approximations are used, putting
\begin{equation}
  \label{Gdrift}
  G_d(u)=u/(2.66 + 1.82u^2),
\end{equation}
\begin{equation}
  \label{lnlambdad}
  \ln\Lambda_d = \ln \left( b_{90d}+ \lambda\over  b_{90d}+1.5r_p\right),
\end{equation}
and
 \begin{equation}
  \label{b90d}
  b_{90d}= Ze\phi_p/[T_i+  \sqrt{100T_iT_er_p/\lde}\; v_f^2/(c_s^2+2.5v_f^2)].
\end{equation}
These are physically motivated and chosen to fit the numerical integrations,
but are not known to be accurate outside the range $10\le
\lde/r_p \le 200$ or for other temperature ratios. 

The total drag force, $F_c+F_o$, is used in section
\ref{colnlessforce} to compare with the drag forces SCEPTIC finds.


\section{Ion Velocity Distribution Effects}
\label{sec3}

\subsection{Plasma Profiles}

\begin{figure}[htp]
  \centering
  \includegraphics[height=3in]{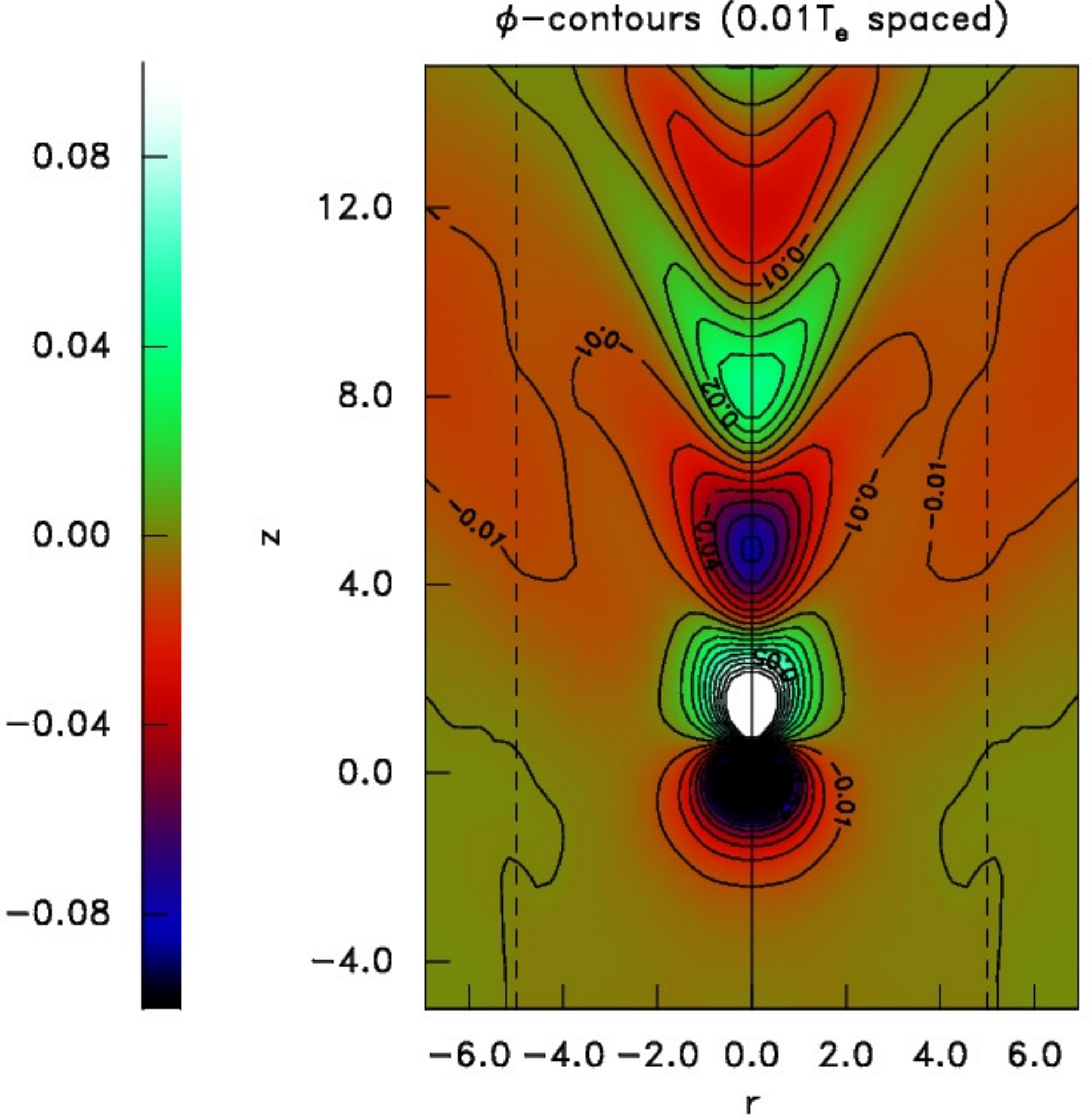}\hskip-2.5in(a)\hskip2.5in 
  \includegraphics[height=3in]{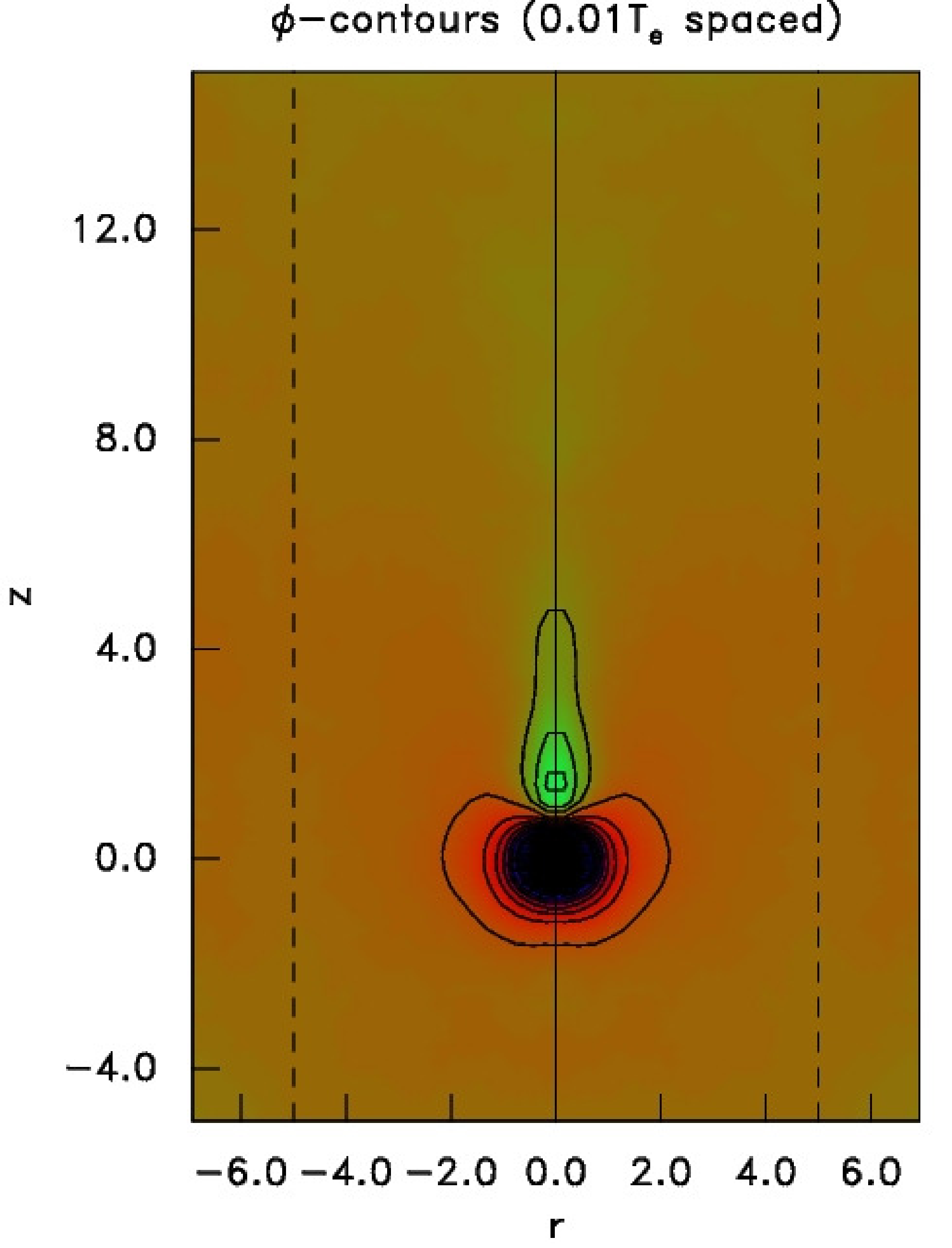}\hskip-2.in(b)\hskip2in\  
  \caption{Wake potential contours $e\phi/T_e$ for $v_f/c_s=1$,
    $r_p=0.1\lde$ $\phi_p=-2T_e/e$, $T_e/T_i=100$, negligible
    collisionality, calculated by COPTIC. Two contrasting background
    ion distributions: (a) shifted Maxwellian; (b) drift distribution
    eq.\ (\ref{eq:driftdist}). Lengths are here scaled to $\lde$ and
    the grain is at the origin.}
  \label{fig:wakeprofiles}
\end{figure}

It is well established that in low collisionality, low $T_i/T_e$,
approximately sound-speed flows the presence of a negatively charged
particle produces an oscillatory wake. Although non-linearity strongly
suppresses the magnitude of the wake potential it does not much change
its form. However, non-linear (and in fact most linear) calculations
of wake structure thus far have used shifted Maxwellian ion
distribution. An example is illustrated in Fig.\
\ref{fig:wakeprofiles}(a), calculated with the COPTIC code, and in
this display averaged over azimuthal angle and over 1000 timesteps.
The first and most dramatic effect of considering collisions is the
replacement of the shifted Maxwellian with the drift distribution eq.\
(\ref{eq:driftdist}). Fig.\ \ref{fig:wakeprofiles}(b) shows the effect
of that replacement. The actual collisional level is completely
negligible ($\nu_c r_p/c_s=10^{-4}$) for all the results of this
section. The two calculations can be
considered collisionless but simply with different background ion
distributions. The drift distribution (b) completely wipes out the
wake oscillations, greatly reduces the height of the trailing peak in
the potential from about 0.2 to 0.03 $T_e/e$, and elongates the peak
further downstream.

\begin{figure}[htp]
  \centering
  \includegraphics[width=3.2in]{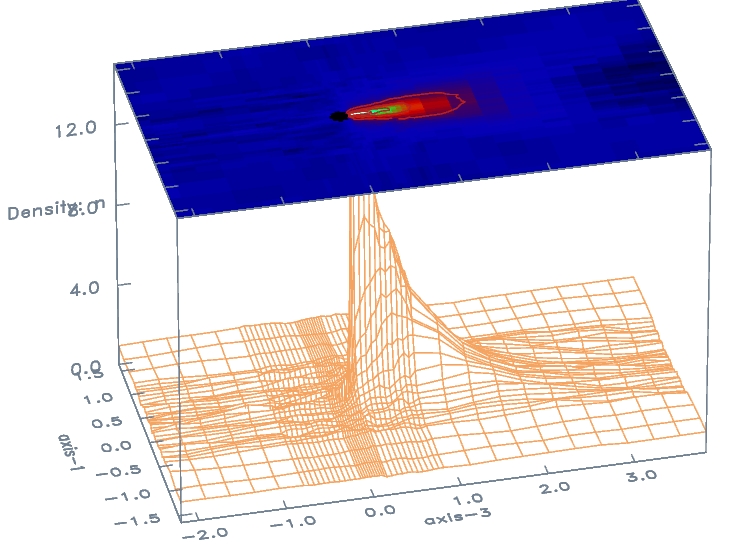}\hskip-.5in(a)\hskip.3in 
  \includegraphics[width=3.2in]{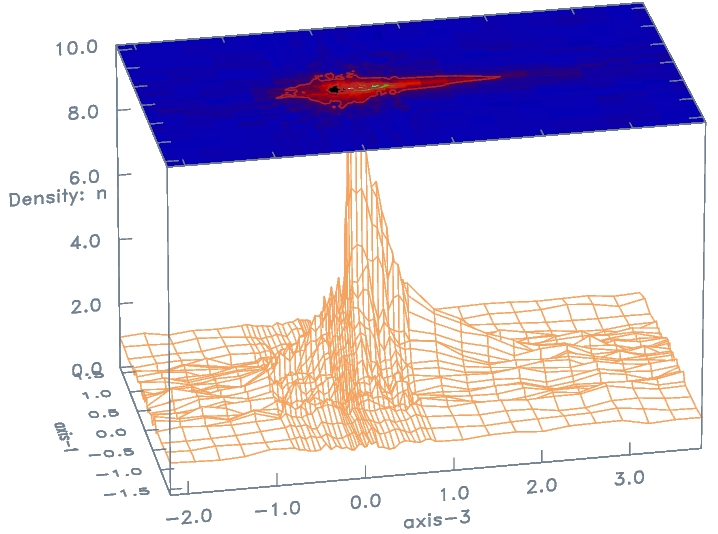}\hskip-.5in(b)\hskip.5in\  
  \caption{Spatial profiles of ion density (scaled to the background) for the
    simulations in Fig.\ \ref{fig:wakeprofiles}. Shift (a), and Drift
    (b) distributions. (A restricted region of the calculation is
    plotted. Axis-3 is $z$, the flow direction. Lengths are scaled to $\lde$.)}
  \label{fig:nprofiles}
\end{figure}

Both of these cases exhibit large ion density enhancements (about 10
times background) trailing the grain (i.e.\ at small positive $z$), as
illustrated in Fig.\ \ref{fig:nprofiles}. These are snap shots of
instantaneous density slices through the three-dimensional domain,
which therefore give an indication of the statistical fluctuations in
the density. Note that the non-uniform grid has much smaller spacing
near the origin to accommodate the fine potential structure
there. Because the grid spacing is smaller than $\lde$, the potential
fluctuations are smaller than the density fluctuations, even before
the averaging performed for Fig.\ \ref{fig:wakeprofiles}.
The predominant difference between the two density profiles is that
the shift distribution (a) has negligible density perturbation
upstream of the grain whereas the drift distribution (b) gives, in
addition to the high peak immediately trailing the grain, a ``halo''
of enhanced ion density around the grain, including in the upstream
direction, and a somewhat longer ridge in the downstream
direction.
 The shift distribution density drops below the background
level at the downstream edge of the region plotted in
Fig.\ \ref{fig:nprofiles}(a), corresponding to the negative oscillation
in potential. The drift distribution (b) does not have this overshoot
in density.

The ``halo'' of ion shielding arises predominantly from the
lower-velocity component of the ion distribution, which is absent from
the shift distribution. Prior simulations of the shift distribution
case\cite{hutchinson06} showed that the upstream shielding cloud is
present for shift velocities lower than approximately $0.5c_s$, and
that the transition in shielding is quite abrupt partly because of
potential asymmetries.  It is therefore of interest to understand
better the ion shielding by ions of low velocity (which are always
present in the drift distribution) and at low drift velocity (which
causes slow ions for the shift distribution).

\begin{figure}[htp]
  \centering
  \includegraphics[height=2.2in]{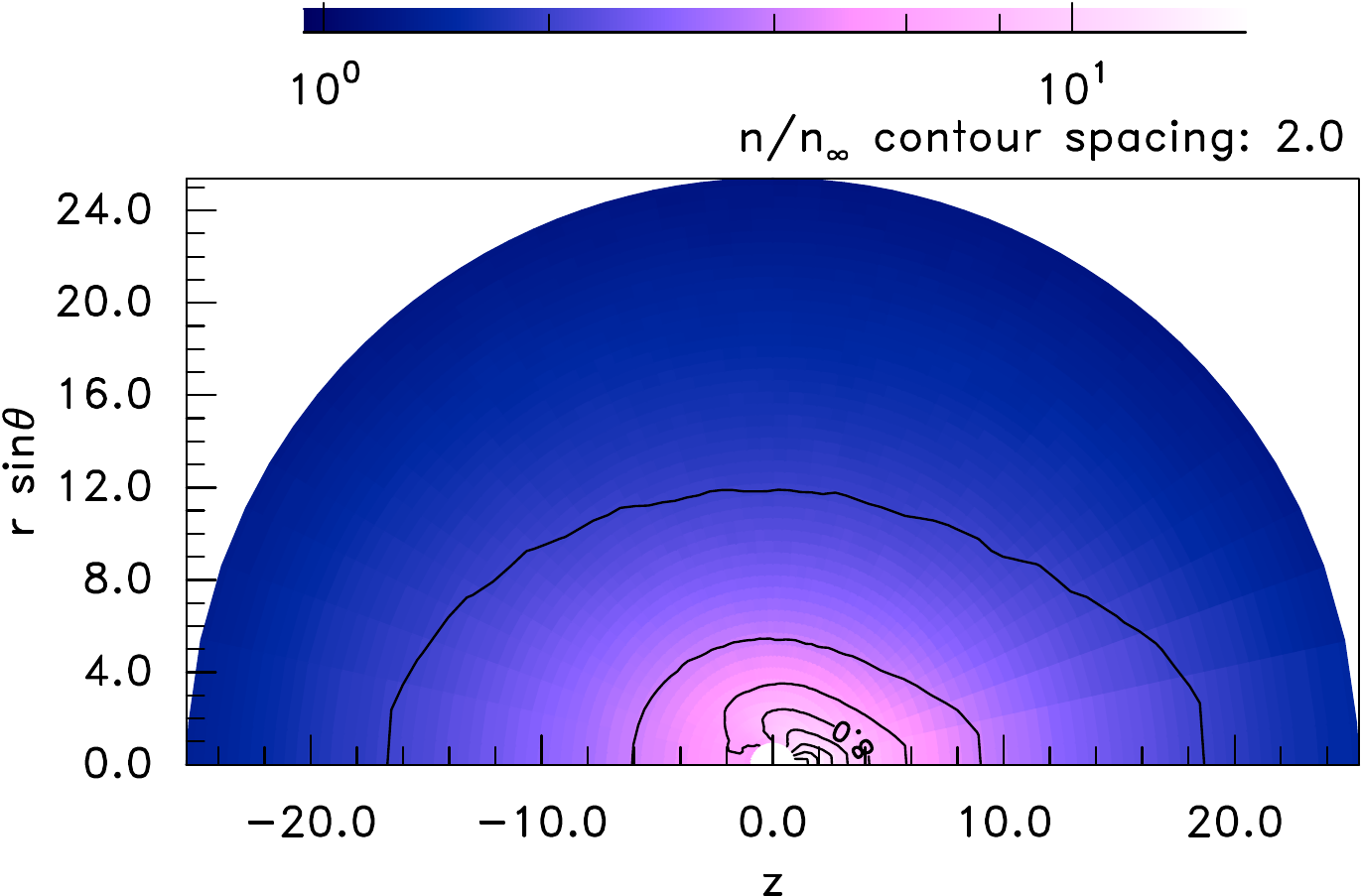}
  \caption{Color contours of density calculated by
    SCEPTIC. $\lde=50r_p$, $v_f=0.2c_s$ drift distribution. Lengths
    are scaled to $r_p$. Flow is in the $+z$ direction.}
  \label{fig:ncont}
\end{figure}
Fig.\ \ref{fig:ncont} shows a contour plot in the $r,z$
plane of the ion density found by SCEPTIC for $v_f=0.2c_s$ with drift
ion-distribution. Here $\lde=50$ (lengths scaled to $r_p$) and only
the inner quarter radius of a total domain $r=100$ is plotted. There
is residual asymmetry in the ion density, consisting of elongation of
the shielding ion cloud both downstream and up, giving both dipole and
quadrupole potential perturbations. In this plot, averaging over 1000
steps has suppressed the cell-to-cell statistical noise level (arising
because the number of ions per cell on this $200\times100$ uniform in
$r$ and $\cos\theta$ spherical grid is only of order unity in the
smallest cells near $r=1$).

\begin{figure}[htp]
  \centering
  \includegraphics[height=3in]{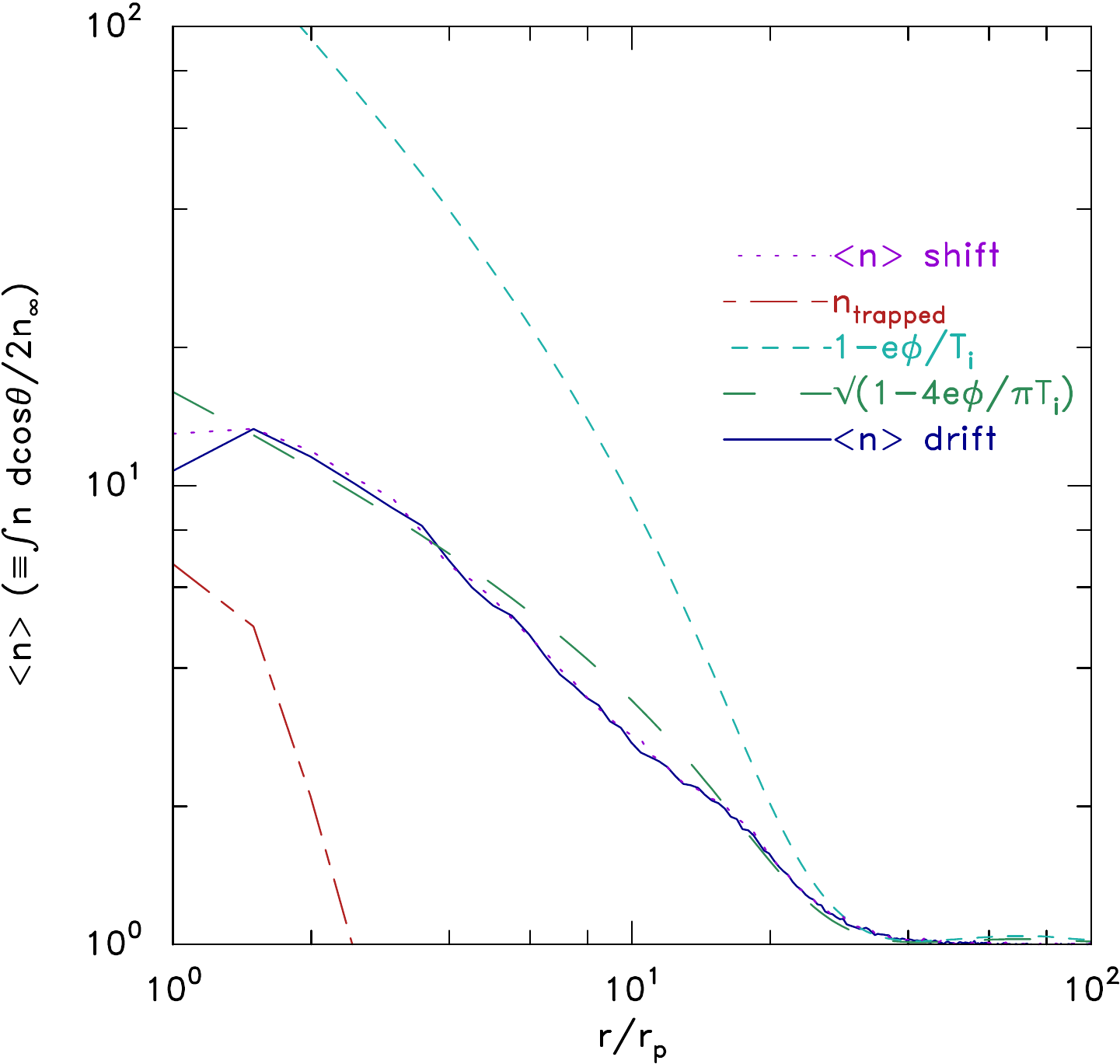}
  \caption{The angle-averaged ion density (solid line) for a SCEPTIC
    calculation with $\lde=50 r_p$, $v_f=0.05c_s$ (drift distribution)
    compared with nonlinear and linear analytic expressions (dashed
    lines). Also shown (dotted) is the shift distribution density,
    virtually the same. Particles trapped in the potential well but
    with sufficient angular momentum to avoid collection are a small
    fraction except very close to the grain.}
  \label{fig:nradial}
\end{figure}
For sufficiently low ion drift (or shift) the density is approximately
spherically symmetric. In Fig.\ \ref{fig:nradial} is shown the
angle-averaged total ion density versus radius and that part of the
density attributable to trapped ions. The solid line is for a
drift case with $v_f=0.05c_s$, but the results are essentially the
same for a shift distribution (shown as dotted line) at this low
$v_f$. For comparison we also show the results of the analytic
approximations for the variation of density with potential ($\phi$,
taken from the angle-average of the simulation). The nonlinear
expression, $n_i/n_{i\infty}= \sqrt{1-4Ze\phi/\pi T_i}$, regarded as a
suitable extension of eq.\ (\ref{eq:attracted}), agrees remarkably
well with the simulation; the linearized expression, $n_i/n_{i\infty}
= 1-Ze\phi/T_i$ is too large by a factor reaching greater than 10.

\begin{figure}[htp]
  \centering
\includegraphics[height=3in]{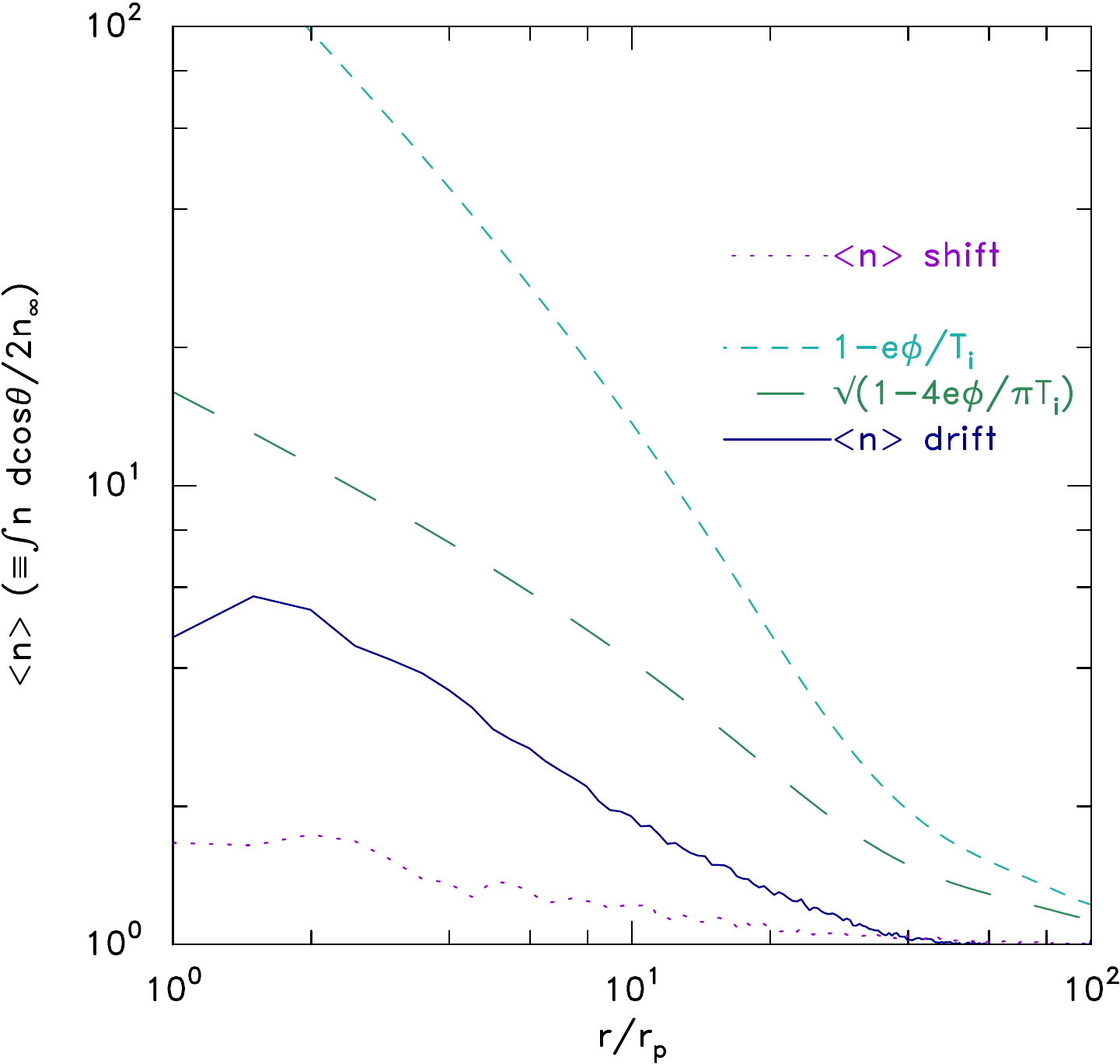}
\caption{The angle-averaged ion density (solid line) for a SCEPTIC
  calculation with $\lde=50 r_p$, but sonic flow $v_f=1.c_s$ with
  drift and shift distributions. The shielding length is much longer
  and the density no longer agrees with the nonlinear symmetric
  expression.}
  \label{fig:nradialv1}
\end{figure}
As the plasma flow is increased, the ion shielding becomes asymmetric,
of course, but also this angle-averaged ion density shows changes for
$v_f\gesim 0.2c_s$. Fig.\ \ref{fig:nradialv1} shows that by the time
$v_f=1.c_s$, the drift ion-distribution gives rise to shielding
density approximately a factor of two less than at low $v_f$, and
correspondingly longer shielding length. However, the shift
distribution shows much greater ion density suppression, so that
ions are hardly shielding the grain at all. There is thus a major
difference in ion shielding between drift and shift distributions,
when $v_f$ is substantial.

\subsection{Collisionless Drag forces}
\label{colnlessforce}

We now document the drag force at negligible collisionality level for
the shift and drift distributions calculated using SCEPTIC, and
compare them with binary collision calculations using Yukawa
cross-section, with the same ion distributions.

\begin{figure}[htp]
  \centering
  \includegraphics[height=2.8in]{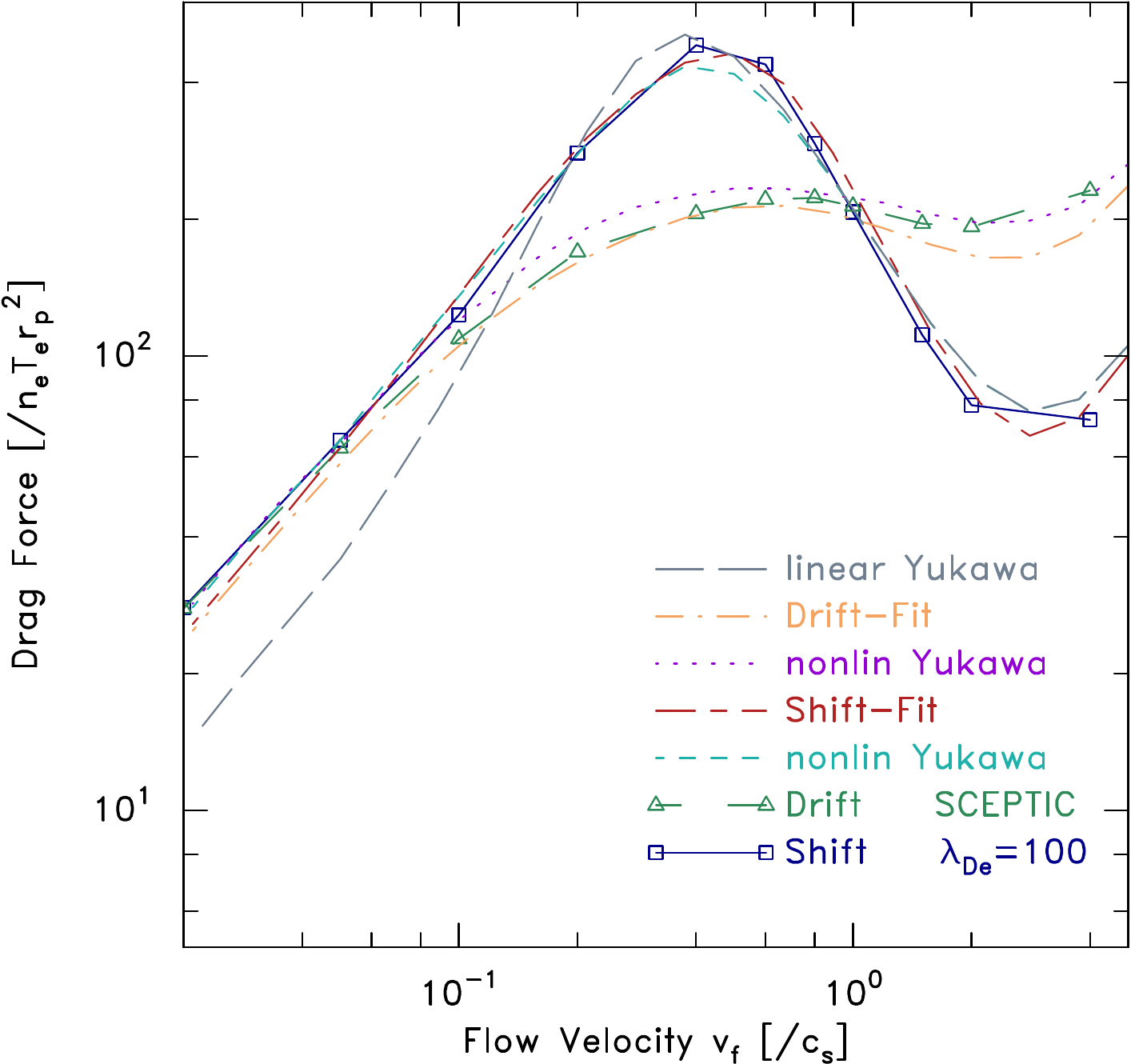}\hskip-2.5in(a)\hskip2.4in 
  \includegraphics[height=2.8in]{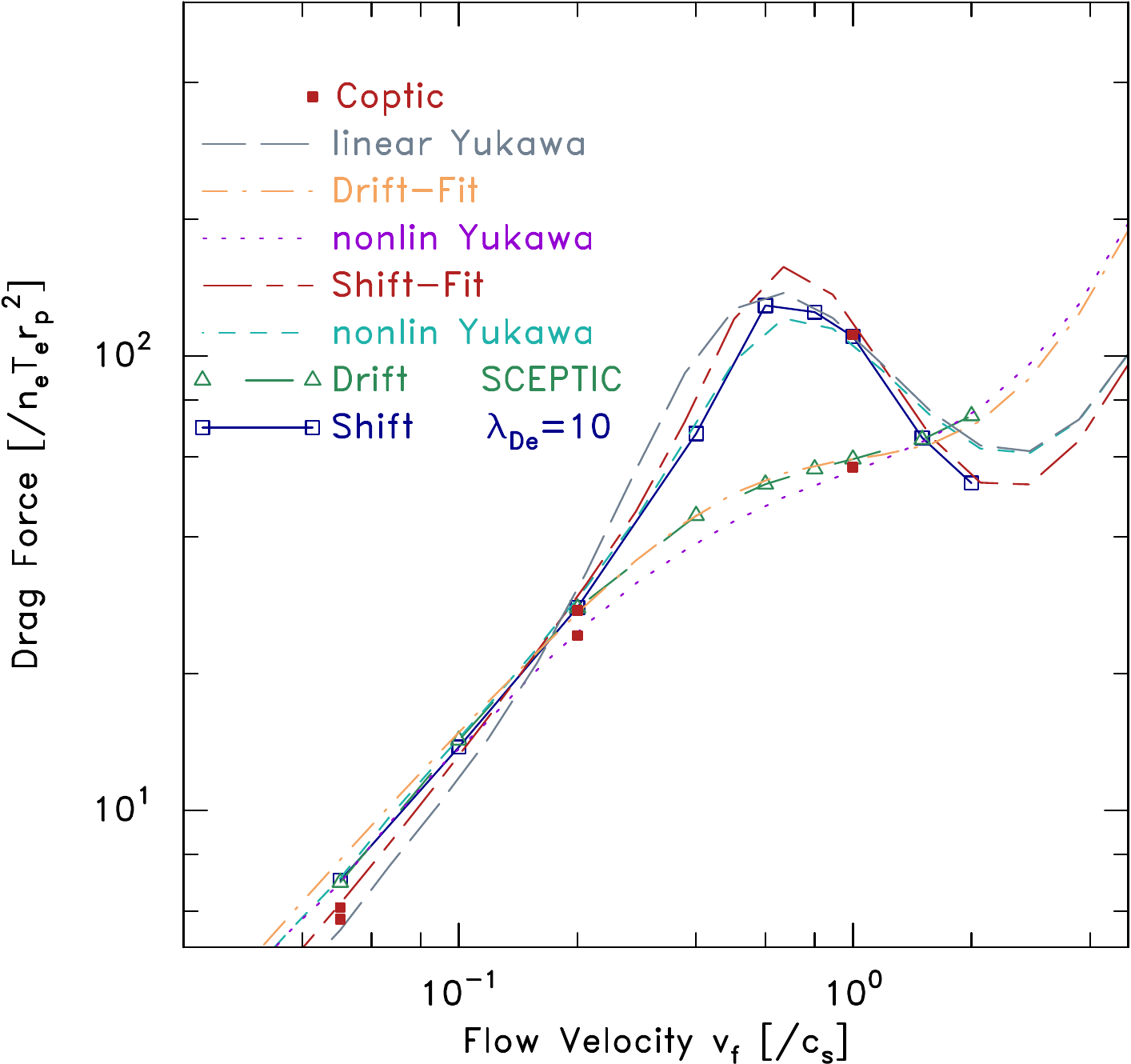}\hskip-2.5in(b)\hskip2.3in\ 
  \caption{Force as a function of ion flow velocity for shift
    ($v_n=v_f$) and drift ($v_n=0$) distributions: (a) $\lde=100r_p$;
    (b) $\lde=10r_p$. Points are SCEPTIC calculations. Short dashed
    lines are the shift-distribution Yukawa cross-section collision
    numerically integrated using nonlinear shielding form (eqs.\
    \ref{adjust},\ref{eq:shiftshield}). Short-long dashes are the
    corresponding analytic Shift-Fit of section
    \ref{orbitalsub}. Dotted and dot-dash lines are the same for the
    drift distribution (eqs.\ \ref{eq:linvd},\ref{eq:shiftshield}).
    Long dashes with short gaps is the shift Yukawa case using
    \emph{linearized} shielding length corrected for flow (eqs.\
    \ref{adjust},\ref{eq:almostlin}). In (b) some independent code
    verification cases from COPTIC are shown.}
  \label{fig:l100nu0}
\end{figure}
In Fig.\ \ref{fig:l100nu0}(a) we see, first, that there is a dramatic
difference in the drag force between the shift and drift distributions
measured by SCEPTIC (points). They agree with each other only at very
small flow velocity $v_f\lesim 0.15c_s$, where the distributions
become almost indistinguishable. For the drift distribution, drag is
approximately constant from $v\sim0.2c_s$ upward. The shifted
Maxwellian, in contrast, has a strong peak at $v\sim 0.5$ more than
twice as high, and then rolls rapidly off to approximately half the
drift-distribution force at $v=2c_s$.  Values derived from integrating
eq.\ (\ref{eq:binaryforce}) using the best nonlinear analytic
estimates (eqs.\
\ref{eq:rcutoff},\ref{adjust},\ref{eq:shiftshield},\ref{eq:linvd}) of
the shielding-length variation, and convenient fits (eq.\
\ref{dragshift} and following, subsection \ref{orbitalsub}) are shown. They agree reasonably well with the
corresponding SCEPTIC results. However, using instead the linearized
shielding length (eq.\ \ref{eq:almostlin}) gives numerically
integrated values that are
nearly a factor of 2 too low at low flow velocity $v\lesim 0.1c_s$).
Fig.\ \ref{fig:l100nu0}(b) shows a similar comparison for ten times
shorter Debye length: $\lde=10$. Reasonable agreement is obtained with
all analytic forms. The discrepancy arising from incorrect shielding
length at low $v_f$ is now mostly compensated by the term $r_p$ in
eq.\ (\ref{eq:almostlin}) for the shift case. Shift and drift
distribution force at $v_f/c_s=0.05,0.2,1.0$ derived from the totally
independent code COPTIC are plotted as filled boxes. They show
agreement with SCEPTIC results within the probable uncertainty of both
codes. The agreement of the Shift-Fit curve is somewhat fortuitous. It
arises because the error induced by using the linear cross-section
approximation eq.\ (\ref{eq:sigma1}) (at $\beta$ values beyond its
applicability) is compensated by an opposite error from using linear
shielding, $\lambda_\ell$, eq.\ (\ref{eq:almostlin}). Of course, that fit was
initially developed using SCEPTIC results, so it is not surprising
that it agrees with the current data.

\begin{figure}[htp]
  \centering
  \includegraphics[height=2.8in]{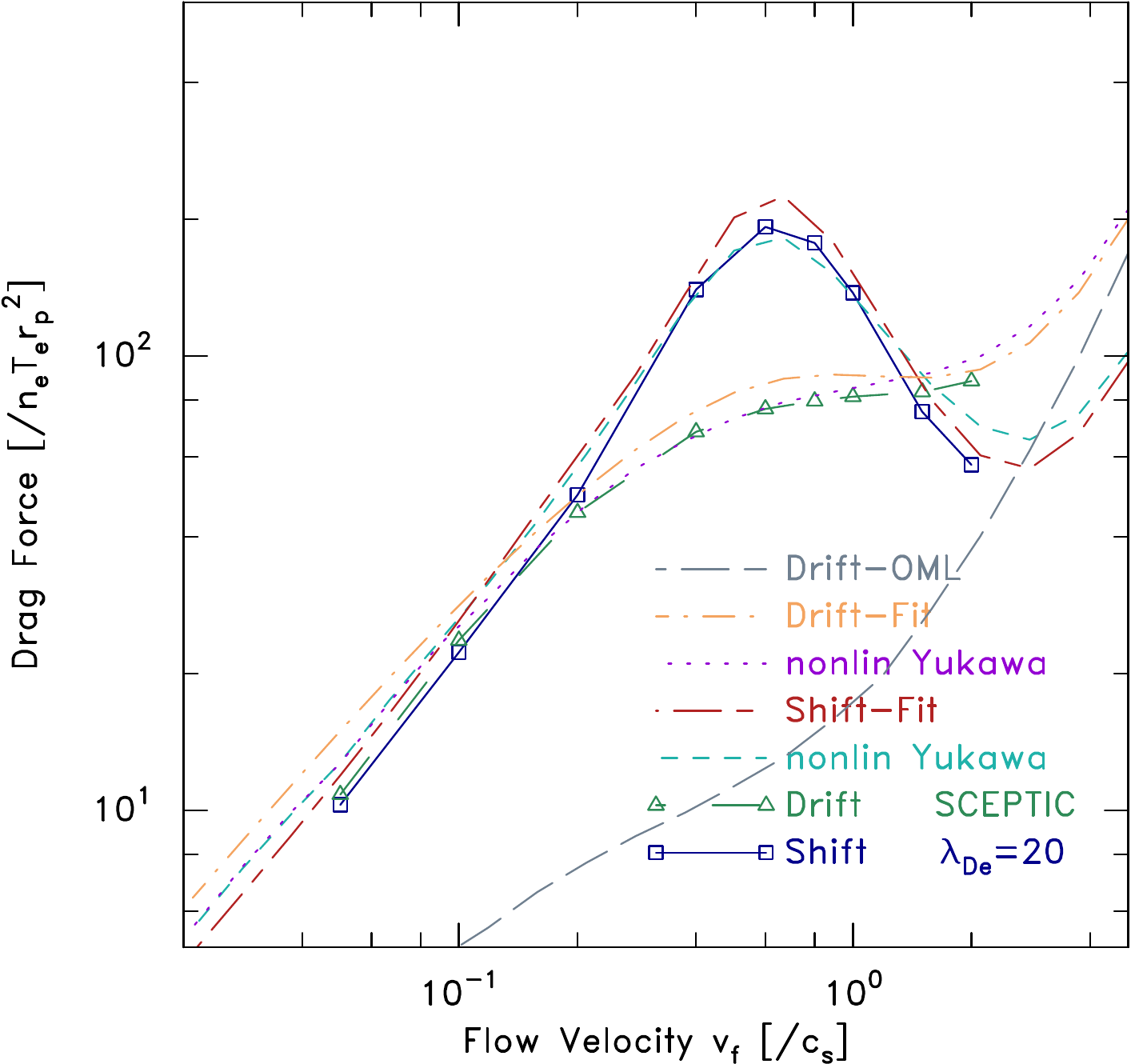}\hskip-2.2in(a)\hskip2.2in 
  \includegraphics[height=2.8in]{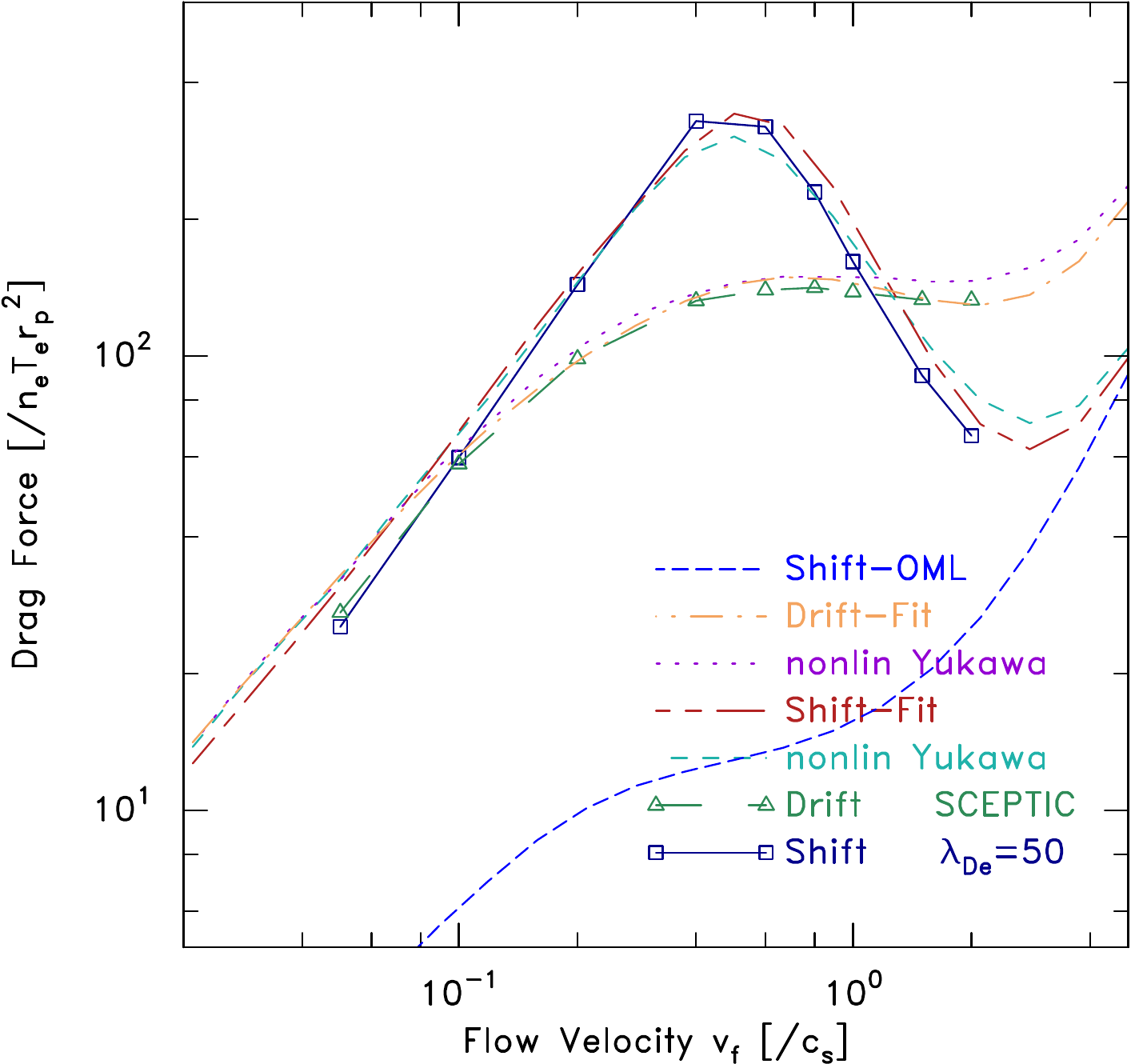}\hskip-2.2in(b)\hskip2.0in\  
  \caption{Like Fig.\ \ref{fig:l100nu0} except that in (a) $\lde=20r_p$ and
    the OML direct collection drift-distribution component is shown;
    in (b) $\lde=50r_p$ and the shift-distribution OML component is shown.}
  \label{fig:l20nu0}
\end{figure}
Fig.\ \ref{fig:l20nu0} shows cases $\lde=20r_p$ and $\lde=50r_p$, in
which the analytic agreement with SCEPTIC results is still quite
good. We also show the direct collection ion component of the forces,
eqs. (\ref{Fcapprox}), (\ref{FcdFinal}), which is included in the
forces of all the plots of these sections, and whose value is
independent of $\lde$. It plays a significant role, especially in the
lower $\lde$ cases, and becomes dominant for all cases when $v_f\gesim
4c_s$.

These observations establish that it is possible to get collisionless
drag force from a Yukawa-potential binary-collision
treatment. However, it is accurate for fairly large $\lde/r_p$ only if
the appropriate shielding length is used. The shielding length value
used must be nonlinear ($r_c$) at low $v_f$, and the variation with
$v_f$ must account for the difference between shift and drift
distributions.

\section{Finite Collisionality Drag Force}
\label{sec4}

We now present a comprehensive documentation of the ion drag force
calculated using SCEPTIC as a function of collisionality, Debye
length, and flow velocity for the drift and shift cases. In all cases
the grain potential is kept fixed at $-2T_e/e$. This might not be the
floating potential, which will undoubtedly vary. However our intention
is to focus on the drag force's intrinsic variation, and for that
purpose it is better not to confuse the issue with another variable
($\phi_p$). The leading dependence of drag is that it is $\propto \phi_p^2$
(see eq.\ \ref{dragshift}).

\begin{figure}[htp]
  \centering
  \includegraphics[height=3in]{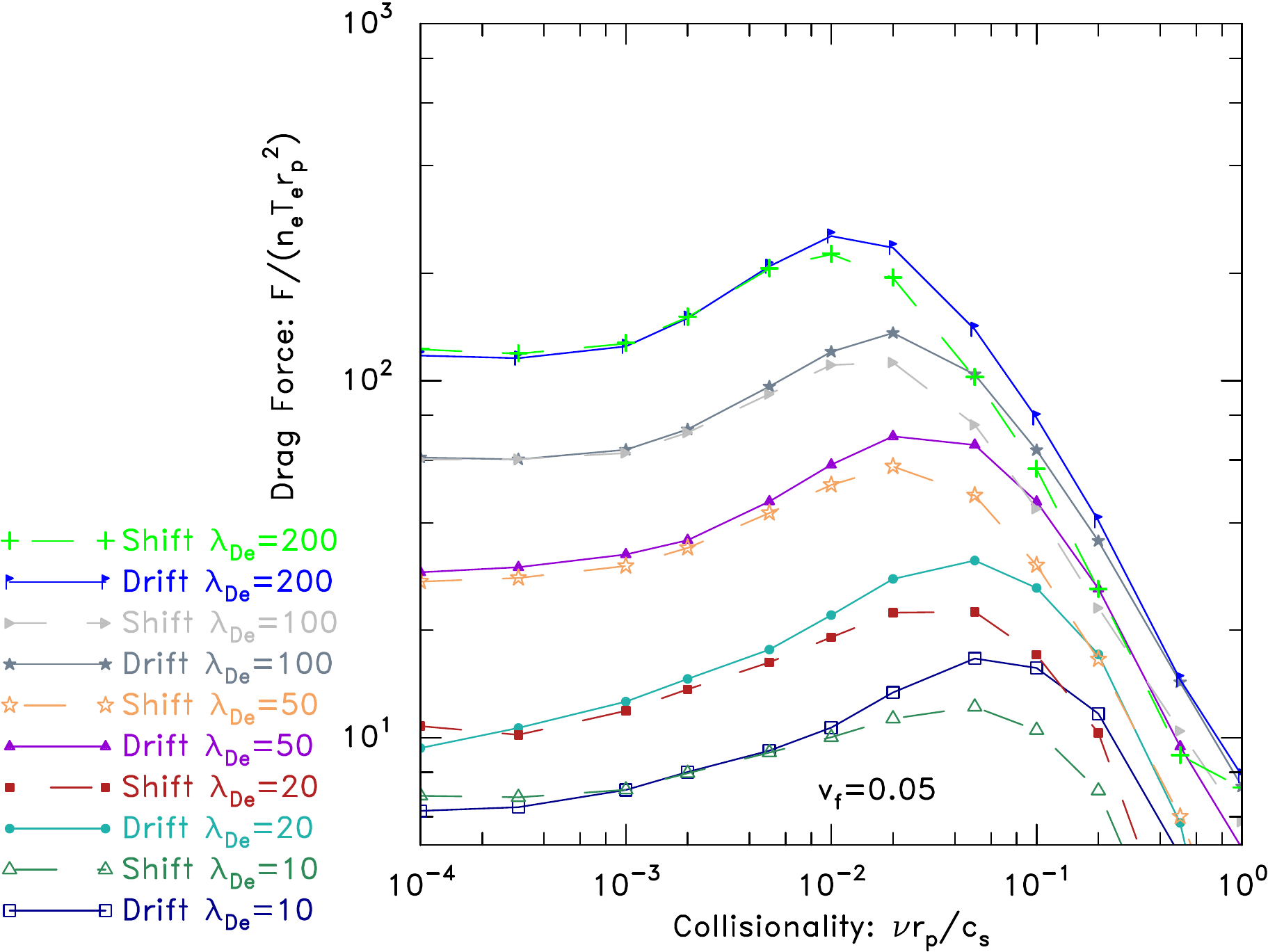}\hskip-2.5in(a)\hskip2.4in\ 

  \includegraphics[height=3in]{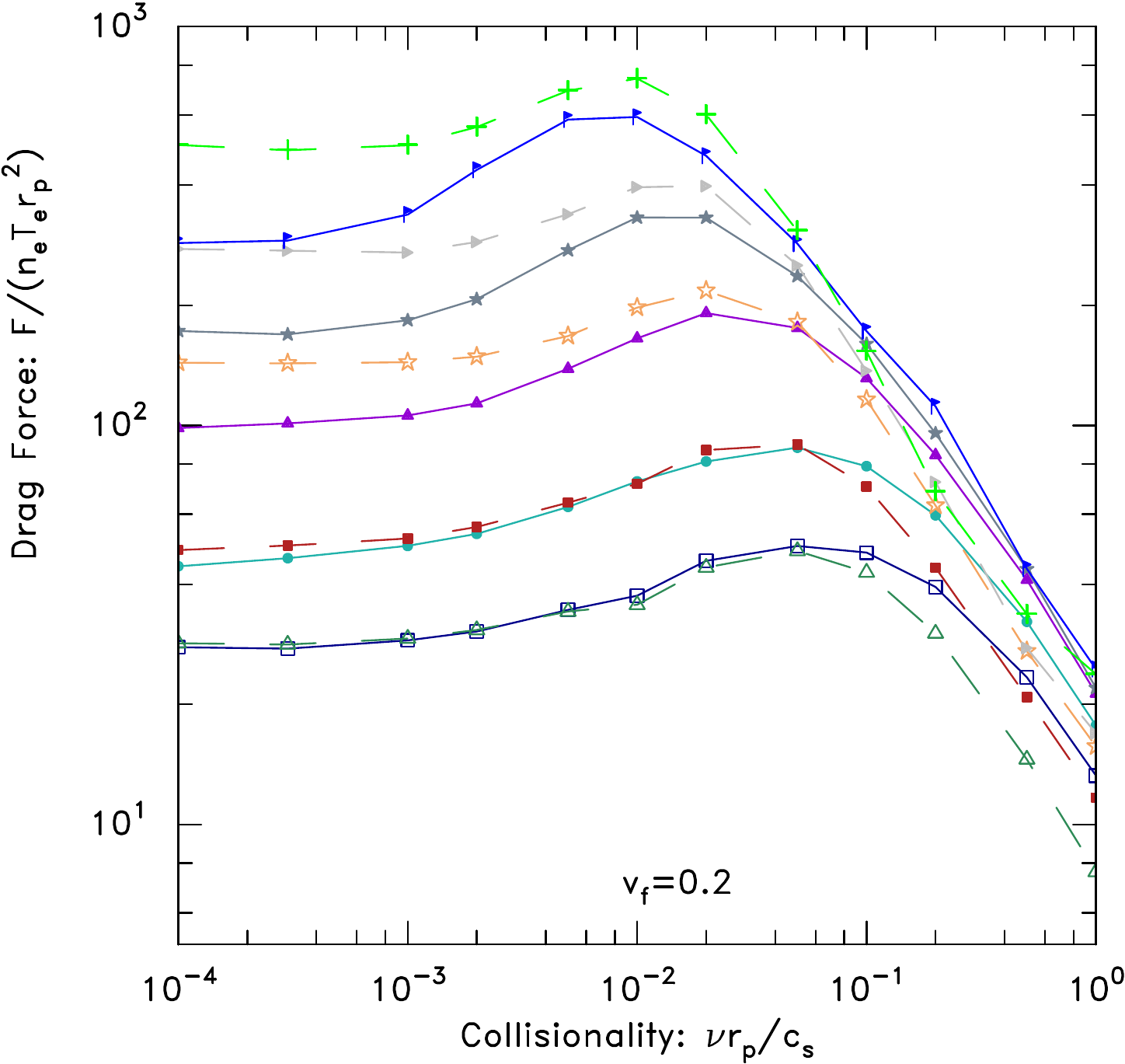}\hskip-2.5in(b)\hskip2.2in\ 
  \includegraphics[height=3in]{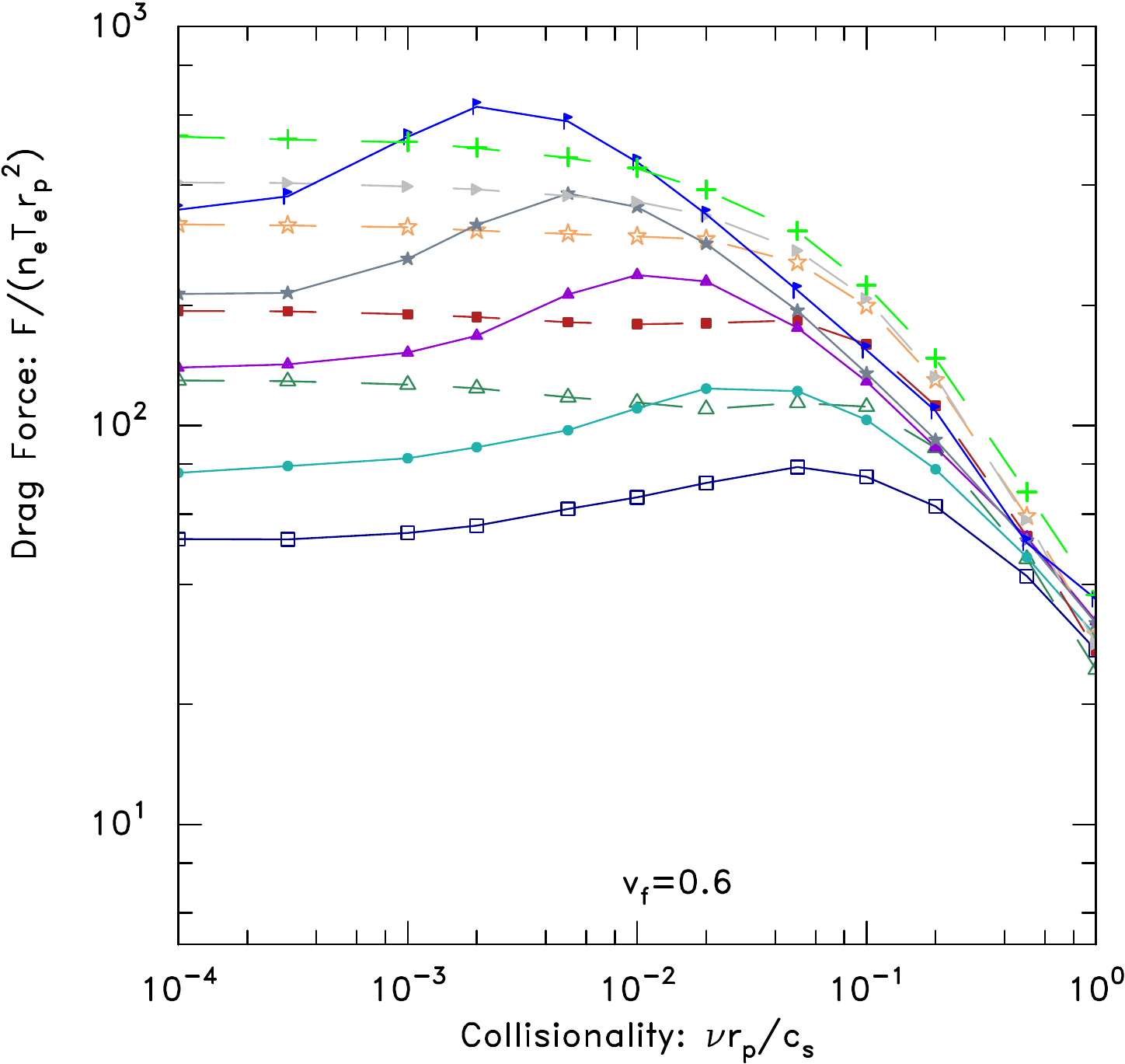}\hskip-2.5in(c)\hskip2.4in\ 
  \caption{Drag force in normalized units as a function of
    collisionality for drift velocity $v_f$ equal to (a) 0.05, (b)
    0.2, (c) 0.6 (times $c_s$). Five different values of Debye length
    (in units of $r_p$) are shown, each for the drift ($v_n=0$) and the
    shift ($v_n=v_f$) case.\label{fvcolnty}}
\end{figure}

Fig.\ \ref{fvcolnty} shows logarithmic plots of drag force variation
with collision frequency $\nu_c$ expressed in units of $c_s/r_p$. We
observe as follows. 

(a) at low flow velocity, $v_f=0.05c_s$, the force at low
collisionality is constant, equal to the zero collisionality values
already given, and the drift and shift cases are the same. As
collisionality increases, the force rises to a maximum roughly a
factor of 2 higher than the collisionless level. Near the maximum, the
drift case force begins to rise above the shift case. Then as
collisionality further increases, the forces begin to decrease
reaching an asymptotic slope of approximately -1 ($F\propto1/\nu_c$).  

(b) at moderate flow velocity, $v_f=0.2c_s$, at smaller $\lde$
similar behavior occurs. But at larger $\lde$ the forces are
unequal at very low collisionality, with the shift case higher than
the drift, consistent with the collisionless behavior given already.
Beyond the peak of the force, the shift case crosses to below the
drift case.

(c) at flow velocity $v_f=0.6c_s$ approaching sonic, the shift case no
longer shows any force increase with collisionality. Instead it stays
almost constant until the high collisionality roll-off is reached. The
drift case still has a peaked shape, but at large $\lde$ it
occurs at lower $\nu_c$ than for the lower flow velocity cases, and
there is a substantial region where the force falls more slowly than
$1/\nu_c$.

In all cases the force is greater for greater $\lde$ when
other factors are equal.

\begin{figure}[htp]
  \centering
  \includegraphics[height=3in]{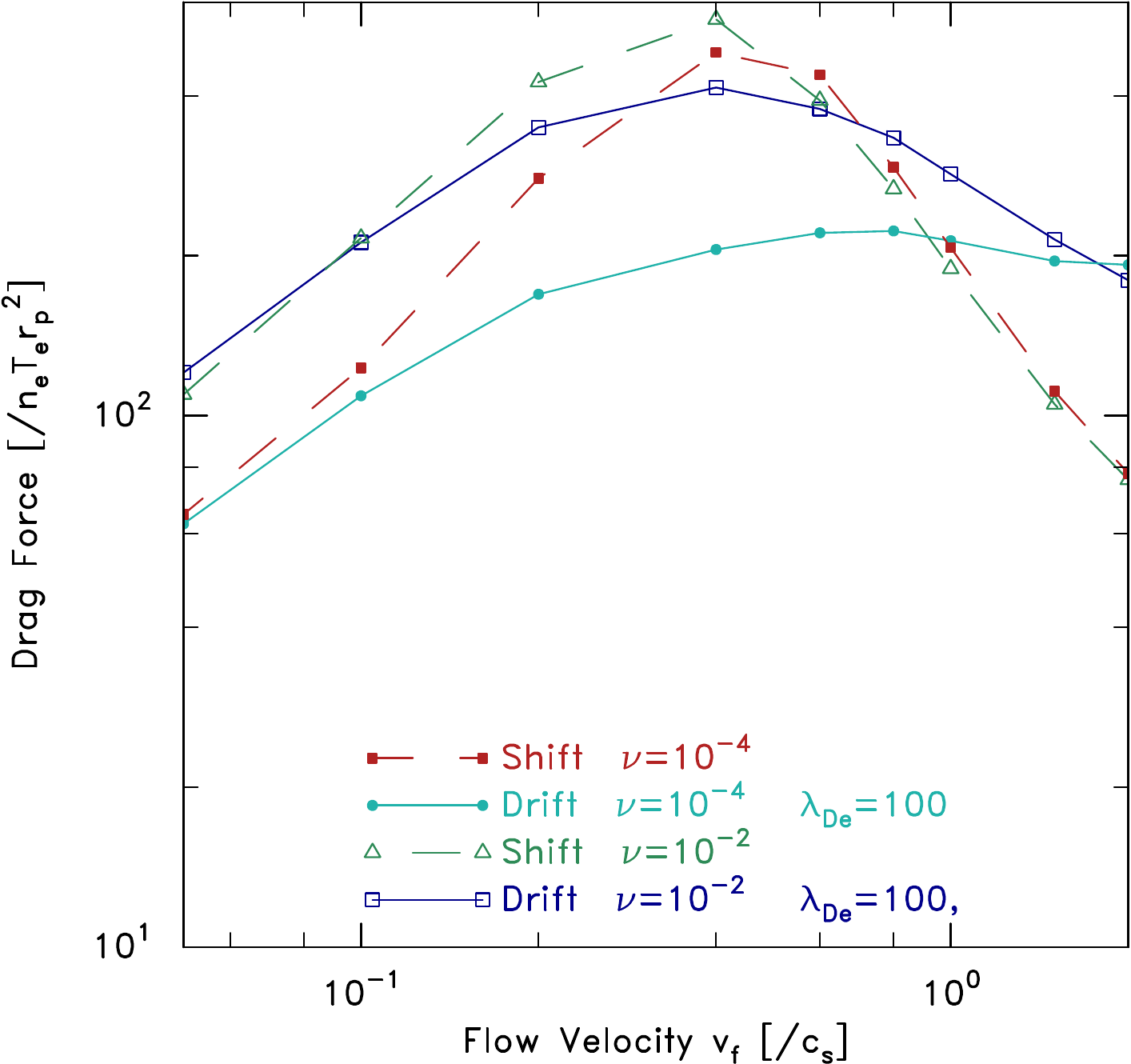}\hskip-2.5in(a)\hskip2.3in\ 
  \includegraphics[height=3in]{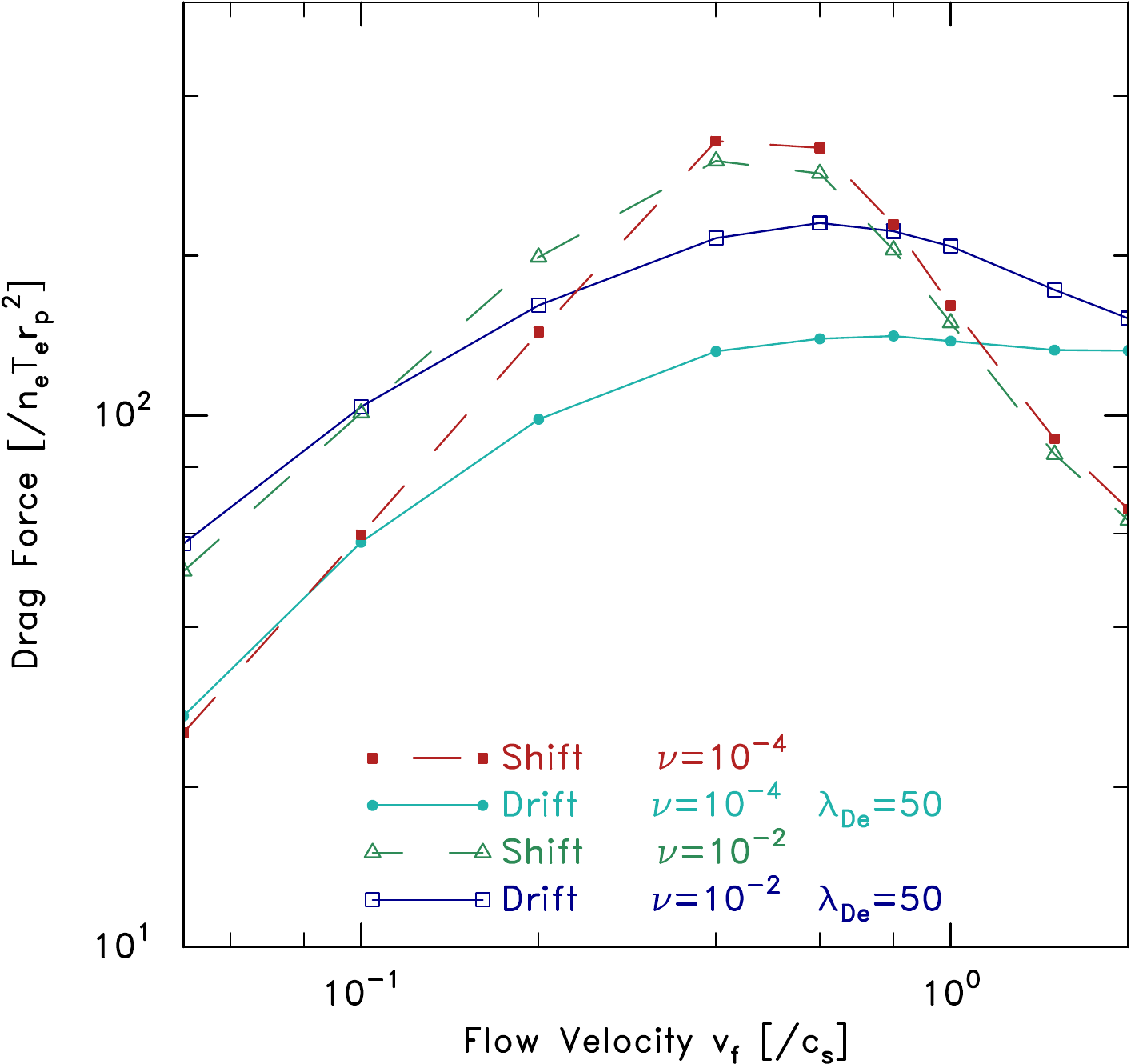}\hskip-2.5in(b)\hskip2.3in\ 

\medskip
  \includegraphics[height=3in]{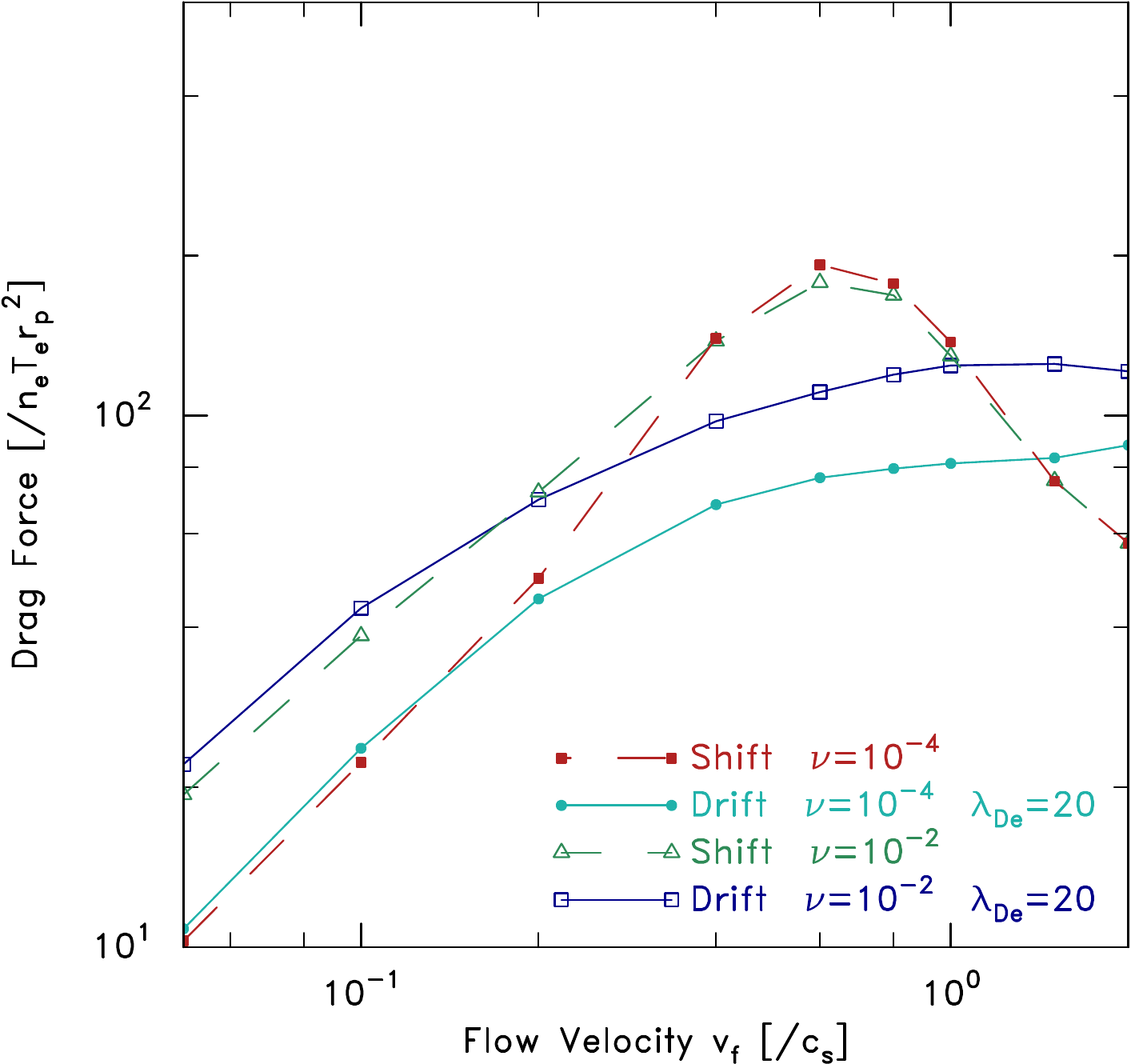}\hskip-2.5in(c)\hskip2.3in\ 
  \includegraphics[height=3in]{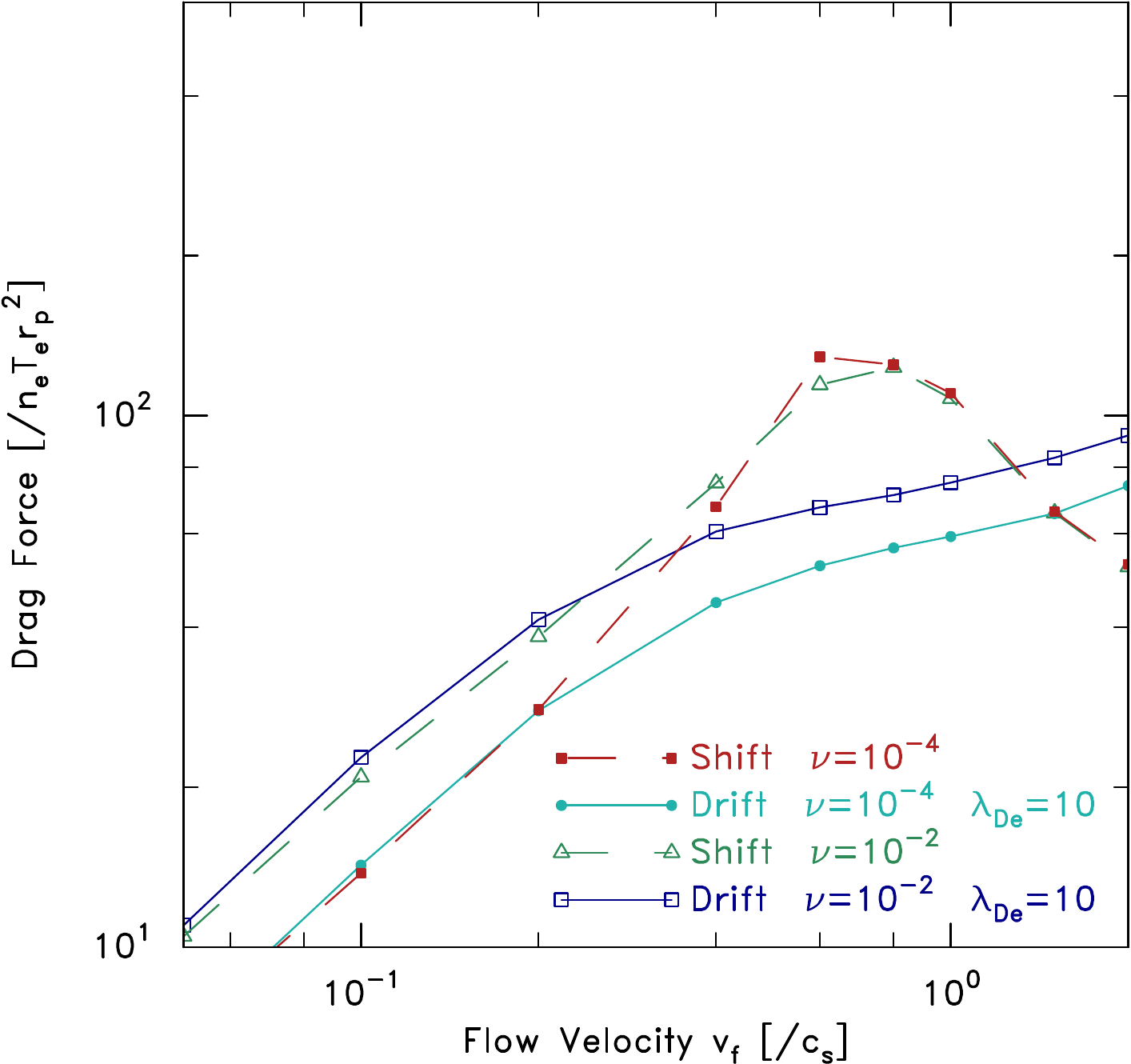}\hskip-2.5in(d)\hskip2.3in\ 
\caption{Drag force as a function of flow velocity for values of the
  Debye length (in units of $r_p$) (a) 100 (b) 50 (c) 20 (d) 10. Two
  values of collisionality $\nu_c$ in units of $c_s/r_p$ are shown, for
  the drift and shift cases. The $\nu_c=10^{-4}$ cases are practically
  collisionless, while the $\nu_c=10^{-2}$ cases are near the peak of
  the collisional force-enhancement.\label{fvvf}}
\end{figure}

In Fig.\ \ref{fvvf} is shown the drag force as a function of flow
velocity. It is therefore a plot along an orthogonal axis of parameter
space. Two values of collisionality are shown: $\nu_c = 10^{-4}c_s/
r_p$, which is essentially collisionless, and $\nu_c = 10^{-2}c_s/r_p$
which is near the peak of the force versus collisionality curves.
We observe the following.

At low flow velocity the drift and shift curves are nearly, but not
exactly the same at $\nu_c = 10^{-2}c_s/r_p$ and both are approximately a
factor of 2 above the $\nu_c = 10^{-4}c_s/r_p$ collisionless values. The
difference decreases somewhat as $\lde$ decreases. The shift
case makes a quite rapid transition so that for $v_f\gesim0.4c_s$
there is no difference between the collisional and collisionless
forces. This transition is, of course, the difference between Figs.\
\ref{fvcolnty}(b) and (c). The Drift case shows no such transition and
the collisional force enhancement is maintained up to $v_f\sim2c_s$
for $\lde= 100$, and further for lower Debye length. The drift
force is far from a $1/v_f$ asymptote for all velocities 
$v_f<2c_s$.
 
To express systematically the modification of the drag force that
arises from finite collisionality, it is best to use a definition of
collisionality that is scaled to the size of the ion shielding cloud,
rather than the grain radius: $\bar{\nu} = \nu_c r_c/c_s$. If we then
divide the drag force by its collisionless value, we get a collisional
\emph{correction factor} $\bar{F}= F(\bar\nu)/F(\bar\nu=0)$ relative
to the collisionless case.

\begin{figure}[htp]
  \centering
  \includegraphics[height=3in]{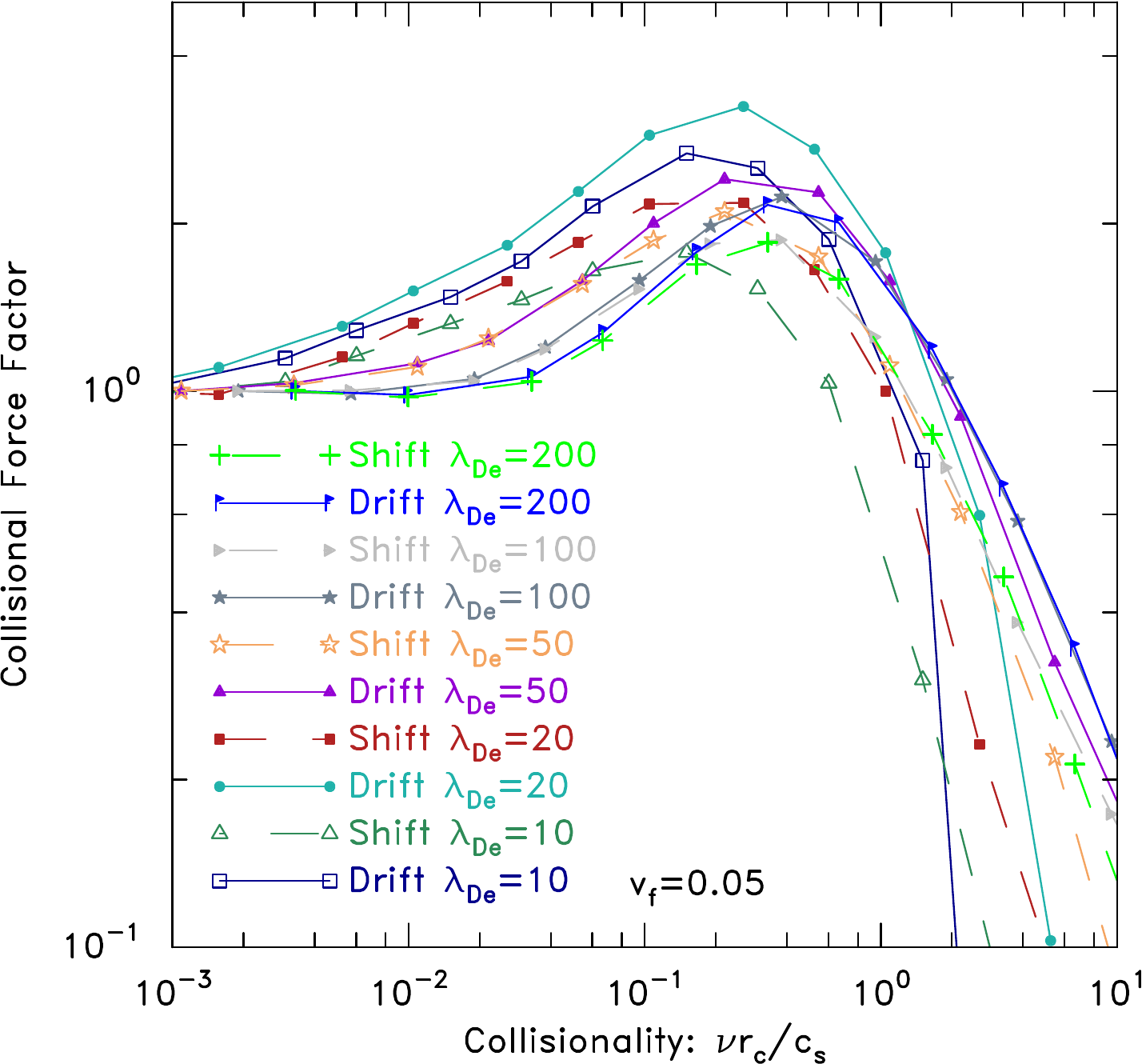}\hskip-2.4in(a)\hskip2.4in\ 

\hbox{\includegraphics[height=3in]{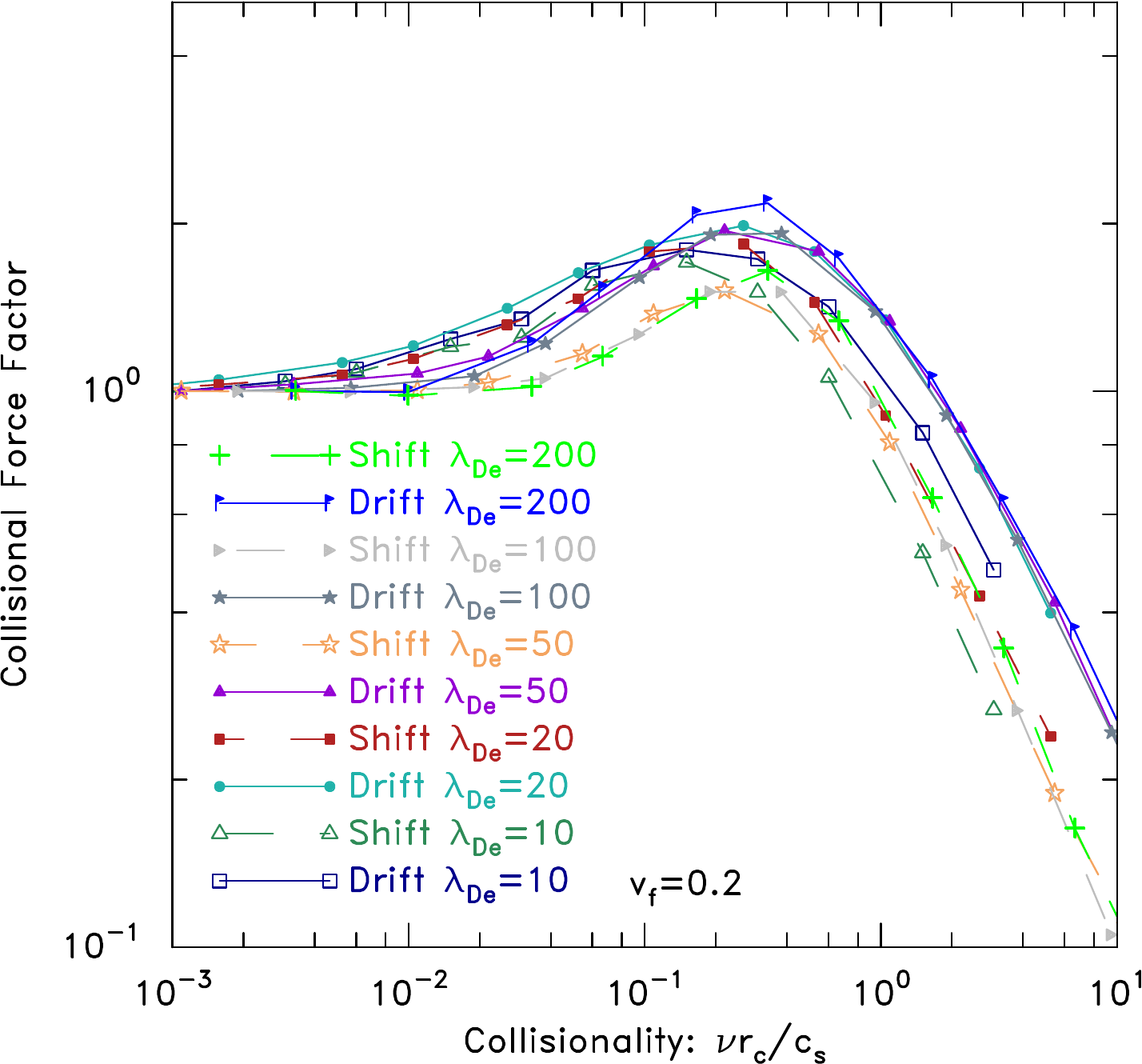}\hskip-2.4in(b)\hskip2.2in
  \includegraphics[height=3in]{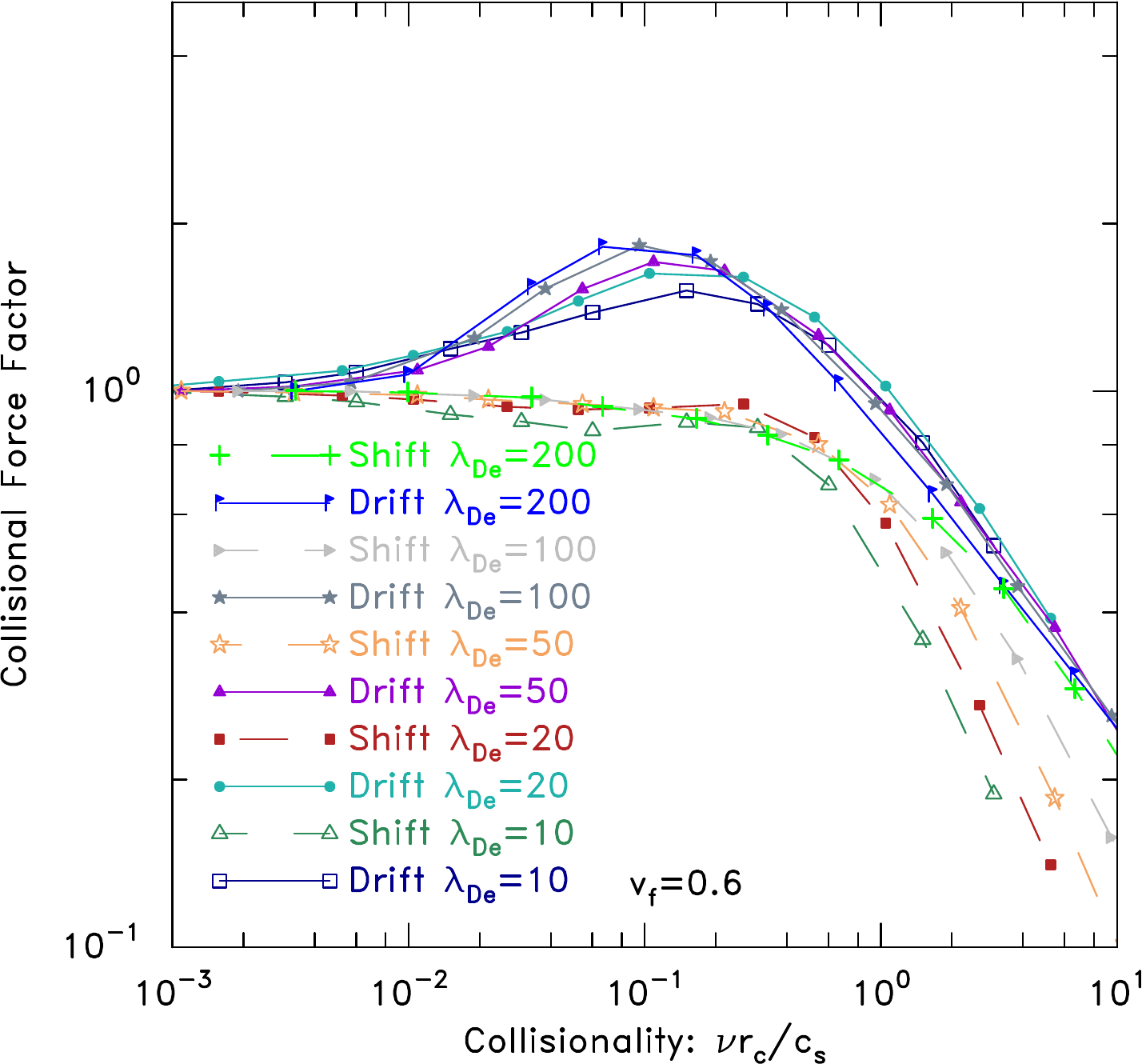}\hskip-2.4in(c)\hskip2.4in\ }
\caption{The drag force divided by its collisionless value,
  i.e. $\bar{F}$, as a function of collisionality $\bar{\nu}=\nu_c
  r_c/c_s$ relative to the ion-cloud radius $r_c$, for drift velocity
  $v_f$ equal to (a) 0.05, (b) 0.2, (c) 0.6 (times $c_s$). Five
  different values of Debye length (in units of $r_p$) are show, each
  for the drift ($v_n=0$) and the shift ($v_n=v_f$)
  case.\label{colfac}}
\end{figure}

Fig.\ \ref{colfac} shows how this scaling reduces the data of Fig.\
\ref{fvcolnty} to approximately universal curves (different for
different velocities) that vary little with Debye length.
The curves can be fitted with a simple rational function:
\begin{equation}
  \label{fcolfac}
  \bar{F}(\bar\nu) = {1 + a \bar\nu \over 1 + b \bar\nu + c \bar\nu^2}
\end{equation}
in which the coefficients are given by Table \ref{fcoltable}.
\begin{table}[htp]
\centering
\begin{tabular}{|l|c|c|c|}
\hline
  Case & $a$ & $b$ & $c$ \\
\hline
 Drift & $(7+30v_f/c_s)$ & $18v_f/c_s$ & $0.5a$ \\
 Shift & $5$ & $8v_f/c_s$ & $3.2$ \\
\hline
\end{tabular}
\caption{Coefficients for the shift and drift cases of eq.\ (\ref{fcolfac}).\label{fcoltable}
}
\end{table}

\begin{figure}[htp]
  \centering
  \includegraphics[height=3in]{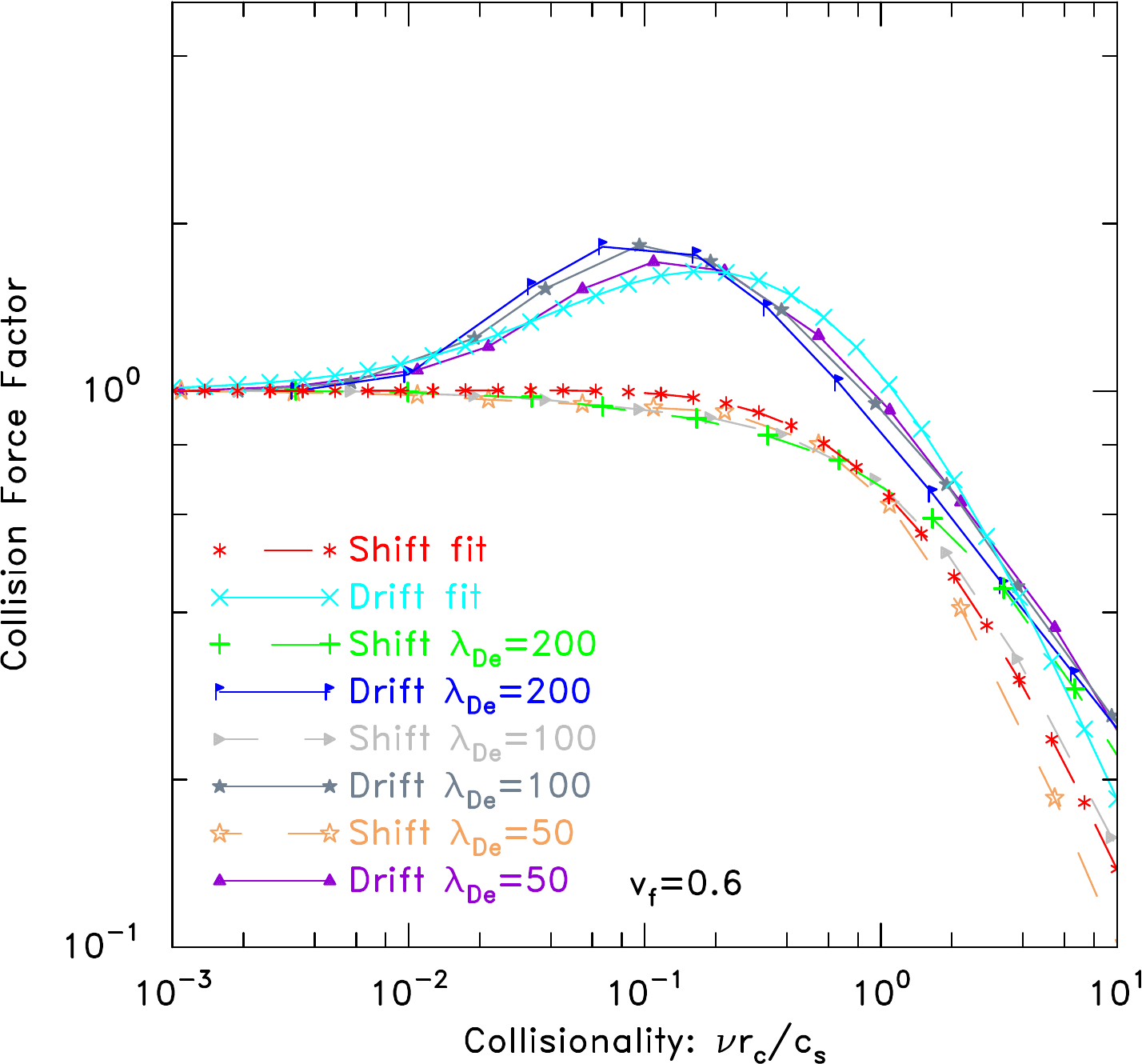}
  \caption{Illustration of fit to collisional force factor, $\bar{F}$,
    from eq.\ (\ref{fcolfac}), for the case $v_f=0.6c_s$.}
  \label{fcolfitfig}
\end{figure}

This fit is optimized for the larger Debye-length cases, which are the
more immediately relevant to most experiments. It can be seen that at
lower Debye length some deviation from universality is present. The
accuracy of the fit may be judged from Fig.\
\ref{fcolfitfig}. However, the fit cannot be trusted for low flow
velocities at collisionalities higher than $\bar\nu \approx 1$. There
the force becomes very small and actually reverses sign.

\section{Discussion}
\label{sec5}

The following qualitative discussion helps to explain the observed
trends. It is based upon the observation that for the large
temperature ratio plasmas under discussion, the plasma shielding of
the grain charge is done mostly by ions. Until the collisionality
becomes strong or the ion flow is fast enough that the electrons
participate strongly in shielding, the shielding cloud consists of a
total charge of ions equal to minus the charge on the grain,
regardless of shielding length. We call these resident ions. Let us
first consider the shift distribution at small flow, $v_f/v_{tn}\lesim
1$.

\subsection{Collisionless Drag Scaling}

Virtually all the ions that enter the shielding cloud and participate
in drag have orbits with large scattering angle, because
$b_{90}>\lambda$. Therefore on average they exit at a random angle
with approximately zero directed momentum. They therefore lose on
average approximately their incoming directed momentum which is
transferred to the grain. This observation has been demonstrated
numerically for Yukawa shielding\cite{Khrapak2003} and is part of the
estimate of the (collisionless) Yukawa momentum cross-section at high
$\beta$.

The residence time of the ions in the cloud is proportional to the
shielding length (i.e. the radius of the cloud) because their speed is
essentially independent of cloud size, and their residency orbit
length is proportional to shielding length.

Therefore the collisionless drag force, which is the rate of momentum
transfer from ions to grain, is proportional to the number of
shielding ions divided by the residence time, and hence the number of
shielding ions divided by the shielding length. When normalized by
density $nT_e r_p^2$, a factor of $\lde^2$ is introduced, which
when divided by shielding length approximately proportional to $\lambda$ gives a
scaling of drag force $\propto \lambda$ in the collisionless
limit. Note that $\lambda$ is not exactly $\propto \lde$
because of the nonlinearity of shielding length, so this is only an
approximate scaling.

\subsection{Effect of a small finite level of collisionality}

Suppose the collision time is longer than the residence time.

Almost all collisions to resident ions produce an ion unable to escape
the grain's potential well, because its prior velocity is replaced by
a velocity $v_i$ drawn from the shifted neutral distribution. If the
shift is small (low $v_f$), then because the ion is on average
resident at a place where $-2\phi/m_i v_i^2$ is $>1$, it is captured.
Some such captured ions will have low enough angular momentum to be
immediately directly collected by the grain. Actually, for smaller
$\lambda_{De}$ a quite large fraction of the collided ions experience
this effect. The average momentum a directly collected ion has after
its last collision before collection, is soon thereafter transferred
to the grain. Any momentum it gains in the field prior to arrival at
the grain is cancelled out by the force it exerts on the grain during
its acceleration. Thus, only the birth momentum, $mv_i$, which has a
directed mean $mv_f$, is gained net by the grain.

The birth momentum of a collision that does \emph{not} lead to
immediate collection is still \emph{on average} transferred during the
ion's trapped phase to the grain by electric force, provided the
trapped ion makes at least one orbit.  Therefore, when the collision
time is longer than the orbit period, regardless of whether a collided
ion is immediately collected or not, its birth momentum is transferred
in a time of order one orbit to the grain.

Whether a resident ion is trapped or not, it still contributes to
shielding, and is still subject to collisions that cause birth
momentum transfer to the grain.  The rate at which collisions are
happening to resident ions is just their number, $N_i=-Q/Ze$, times
the collision frequency $\nu_c$.  There is therefore an additional
drag force, arising from collisions, approximately equal to
\begin{equation}
  \label{colnfrc}
-Q\nu_c mv_f/Ze.  
\end{equation}
It is \emph{independent} of shielding cloud size and proportional to
collision frequency and flow velocity.

Prior to a collision, the resident ion has transferred some of its
incoming momentum (or if it previously collided, its birth momentum)
to the grain. The presumption, so far, is that provided the collision
time is longer than the residence time, on average all of the
incoming momentum has already been transferred, by the time a
collision occurs. In the nonlinear regime under consideration, most
resident orbits in a Yukawa potential in fact have greater than 180
degree scatterings, and many more than 360 degrees \cite{Khrapak2003}.
Since rare collisions occur on average half way through the residency,
approximately 180 degree residency implies on average 90 degree
scattering prior to a collision, and hence full momentum transfer,
justifying the presumption.

\subsection{Collision time comparable to residence time}

If the collision time becomes shorter than the time of scattering
through 180 degrees, then the momentum transfer prior to the first
collision will average less. However, in that case, multiple
collisions will be important too, and more complicated considerations
arise. At the threshold of onset of these complications, an ion has on
average approximately one collision during a transit. That is, the
collision time is equal to the residency time. If that is the case,
then the drag will have been increased by a factor of approximately
two, because a transiting ion will have transferred its incoming
momentum plus one collision's worth of birth momentum (both of which
are on average $mv_f$) to the grain. Therefore the collisional effects
cannot increase the drag by more than approximately a factor of two
before multiple collision effects have to be considered.


If the collision time is noticeably shorter than the orbit time, then:
(1) The incoming momentum is not all transferred to the grain.  (2)
The birth momentum of each collision is also not all transferred to
the grain.  Collisions intervene during the momentum transfer process
and transfer what remains to the neutrals instead.

\subsection{Drift case}

In the drift case at long collision time, what happens is that a
collided ion is (re)born with zero average momentum.  There is
therefore no momentum transfer per se arising from collisions.
However, all resident ions are instead acted upon by the accelerating drift
force field $D=\nu_c mv_f$. Resident ions transfer essentially all of this
momentum to the grain by virtue of the potential well electric
field. So collisions introduce an additional drag force that is
simply $N_i D = N_i\nu_c mv_f$. This is the same as the shift additional
collisional force, eq.\ (\ref{colnfrc}). 

The collisional enhancement becomes comparable to the collisionless
drag term (only) when the collision time becomes comparable to the
residence time. When the collision time is shorter than the residence
time, not all the momentum gained from $D$ is transferred to the
grain.  Therefore virtually the same argument applies to the Drift
case as the shift case. The drag enhancement due to collisions
cannot exceed approximately a factor of 2.

\subsection{Large flow velocity}

When $v_f$ becomes $\gg v_{ti}$, there is a major difference between the
drift and the shift case. 

Drift ions are still born with the same, zero-average, velocity. There
is essentially no difference in the argument above, the collisional
force can be considered to be just the driving force field acting on
the shielding cloud.

Shift ions, however, are born with an average velocity equal to the
flow velocity. If this is large enough to exceed the potential-well
escape velocity (from their birth position), then they \emph{will not}
transfer all their momentum to the grain, regardless of
collisionality. Instead they will transfer only the momentum
corresponding to their energy loss escaping the well, and this will be
a small fraction once the birth energy substantially exceeds the
(average) potential depth. This explains why the collisional force
enhancement disappears for the shift case when $v_f \gesim 0.3c_s$ or
a bit lower at higher $\lde$. This is the velocity where the flow
energy exceeds roughly $\phi_p/2\lambda$, and birth ions are no
longer trapped.

\subsection{Continuum Regime}

When the collision time becomes substantially shorter than the orbit
time, we begin to move into the continuum regime where collisions
dominate. We find that the charge on the grain is no longer completely
shielded. Residual Coulomb-like electric field is required to attract
the ions from large distances into the potential well and eventually
to be collected\cite{patacchini09}. In this regime the drag begins to
fall off approximately inversely proportional to collisionality, but
under some circumstances actually reverses sign. It has been
demonstrated\cite{Patacchini2008} that SCEPTIC force calculations
agree with analytic solutions to the continuum equations at high
collisionality.

An unexpected new observation in the present data at intermediate and
high collisionality is that at low flow velocity, the ion drag force
for the drift case significantly exceeds the shift case. The SCEPTIC
results show that drag in these cases arises from an enhancement of
the ion density downstream of the grain, caused by focusing of the
ions as they flow past.  The drag \emph{difference} is observed in the
simulations to be a result of the downstream density enhancement being
greater for the drift case. The direct ion momentum collected by the
grains in at low flow velocity (e.g.\ $v_f=0.05c_s$) is actually a
negative contribution to the drag, but until one is deep into the
collisional regime the total drag is dominated by the electric field
force that arises from the downstream density enhancement. The small
($\sim 10\%$) density asymmetry gives rise to an even smaller ($<1\%$)
potential deviation from spherical symmetry. Yet this is what is
responsible for the drag; and the asymmetry magnitude is different for
the drift and shift cases. We speculate that the density difference
can be explained heuristically as arising from the combination of
non-uniform accelerating field (approximately Coulomb field close to
the grain) and finite Knudsen number (collision mean free path
relative to scale size), which can be shown to lead to a difference in
flow velocity between drift and shift cases.

\section{Summary}

The ion drag force on a spherical grain in a flowing collisional
plasma has been calculated using the SCEPTIC and COPTIC particle in
cell codes over a wide range of collisionality, flow velocity, and
Debye length, for grain potential $-2T_e/e$ and temperature ratio
$T_e/T_i=100$. These self-consistent calculations take into account
the fully non-linear behavior that arises with typical dusty plasma
experiments because the ion ninety-degree impact-parameter length
(sometimes called the Coulomb radius) is similar to or exceeds the
Debye length.

For negligible collisionality, it is found that using the ion
distribution function appropriate for \emph{drift} driven by a force
field makes a major difference compared with using a \emph{shift}ed
Maxwellian in which the same ion flow arises from neutral background
flow without a force field. In comparison with the more widely studied
shifted Maxwellian case, the wake potential of the drift case is far
smaller in magnitude and has no oscillations (regardless of
collisionality). The low-collisionality drift-distribution drag force
is up to a factor of two smaller than the shift-distribution force in
the subsonic flow range ($0.2<v_f/c_s<1$.) and up to a factor of two
larger for supersonic flow. These differences can be explained on the
basis of approximating the ion interaction with the grain as being
scattering in a Yukawa potential. However, quantitative agreement is
obtained only if the full nonlinearity not only of the scattering
cross-section, but also of the shielding length is accounted
for. Comprehensive analytic formulas representing the integrated
forces, and verified in comparison with the simulations, are
given\footnote{See supplementary material at [URL to be inserted by
  AIP] which provides computer routines for force evaluation.}.

Finite collisionality initially enhances the drag force, but only by
up to a factor of $\sim 2$ relative to the negligible collisionality
value. At flow velocities greater than $\sim 0.5 c_s$, collisional
drag enhancement occurs only for drift distributions, not shift
distributions.  As collisionality increases further, with collision
frequency above $\sim0.2c_s/r_c$, the drag force
falls off approximately inversely with collisionality (but
eventually reverses). Surprisingly, it is observed that at very low
flow velocity, where the \emph{collisionless} drag for drift and shift
distributions is the same, the \emph{collisional} drag force for the
drift distribution exceeds that for the shift distribution by roughly
a factor of 2.

The collisional drag force enhancement can be represented by an almost
universal function of scaled collisionality and flow velocity, for
which simple fits are provided.


\begin{acknowledgments}
  Work supported in part by NSF/DOE Grant DE-FG02-06ER54982. C B
  Haakonsen was supported in part by the Department of Energy Office
  of Science Graduate Fellowship Program (DOE SCGF), administered by
  ORISE-ORAU under contract no. DE-AC05-06OR23100. Some of the computer
  simulations were carried out on the MIT PSFC parallel AMD
  Opteron/Infiniband cluster Loki, concerning which we thank John
  Wright and Darin Ernst for advice.
\end{acknowledgments}
\bibliography{mybib}

\end{document}